\documentclass[11pt]{article}
\usepackage{amssymb}
\usepackage{latexsym}
\usepackage[margin=1.15in]{geometry}

\usepackage{mathptmx}

\title{Rapidly Mixing Markov Chains: A Comparison of Techniques \\
{\large (A Survey)}}

\author{Venkatesan Guruswami\thanks{Computer Science Department, Carnegie Mellon University, Pittsburgh, PA. {\tt guruswami@cmu.edu}. Survey written in 2000 when the author was a student at MIT.}}

\date{{\bf Disclaimer:} This unpublished survey was written in 2000, and is being posted \emph{unedited in its original form}, in response to requests for a permanent URL that can be cited. \\ It is thus \emph{outdated} and does {\em not} reflect the state of the art in 2016.}

\newcommand{\real}{{\mathbb R}}
\newcommand{\mcmc}{Markov chain Monte Carlo}
\newcommand{\M}{{\mathfrak M}}
\renewcommand{\P}{{\mathfrak P}}
\newcommand{\res}{{\mathfrak R}}
\newcommand{\xor}{\oplus}
\newcommand{\kap}{\kappa}
\newcommand{\ra}{\rightarrow}
\newcommand{\baup}{balanced almost uniform permutations}
\newcommand{\dprod}[2]{\langle #1,#2 \rangle}

\renewcommand{\bot}{\it Bot}
\renewcommand{\top}{\it Top}

\def\pr{\mathop{\bf Pr}\limits}
\def\av{\mathop{\bf E}\limits}

\newcommand{\eps}{\varepsilon}

\newcommand{\proof}{\noindent {\bf Proof:}\hspace{1em}}
\newtheorem{theorem}{Theorem}[section]
\newtheorem{lemma}[theorem]{Lemma}
\newtheorem{prop}[theorem]{Propostion}
\newtheorem{cor}[theorem]{Corollary}
 
\newtheorem{defn}{Definition}[section]

\newcommand{\eqdef}{{\stackrel{{\rm def}}=}}

\begin{document}
\maketitle
\thispagestyle{empty}

\vspace{-4ex}
\begin{abstract}
For many fundamental sampling problems, the best, and often the only
known, approach to solving them is to take a long enough random walk
on a certain Markov chain and then return the current state of the
chain.  Techniques to prove how long ``long enough'' is, i.e., the
number of steps in the chain one needs to take in order to be
sufficiently close to the stationary distribution of the chain, are of
great importance in obtaining estimates of running times of such
sampling algorithms.

In this report, we survey existing techniques to bound the mixing time
of Markov chains. The mixing time of a Markov chain is exactly
captured by the ``spectral gap'' of its underlying transition
matrix. The spectral gap is closely related to a geometric parameter
called ``conductance'' which is a measure of the ``edge-expansion'' of
the Markov chain. Conductance also captures the mixing time up to
square factors. Lower bounds on conductance, which give upper bounds
on the mixing time, are typically obtained by a technique called
``canonical paths'' where the idea is to find a set of paths, one
between every unequal source-destination pair, such that no edge is
very heavily congested.

Unlike conductance, the canonical paths approach cannot always show
rapid mixing of a rapidly mixing chain. It is known that this
``drawback'' disappears if we allow the flow between a pair of states
to be spread along multiple paths.  We prove that for a large class of
Markov chains, including all the ones that we use in the sampling
applications we will be interested in, canonical paths does capture
rapid mixing, i.e., we show that small mixing time implies the
existence of some collection of paths with low edge congestion.
Allowing multiple paths to route the flow still does help a great deal
in the design of such flows, and this is best illustrated by a recent
result of Morris and Sinclair~\cite{MS99} on the rapid mixing of a
natural Markov chain for sampling $0$-$1$ knapsack solutions; this
result seems to rely critically on fractional flows.

An entirely different approach to prove rapid mixing, which in fact
historically preceded the conductance/canonical paths based approach,
is ``Coupling''. Coupling is a very elegant technique and has been
used to prove rapid mixing of several chains where designing good
canonical paths seems to be a hideous task. ``Path Coupling'' is a
related technique discovered by Bubley and Dyer~\cite{BD97} that often
tremendously reduces the complexity of designing good Couplings. We
present several applications of Path Coupling in proofs of rapid
mixing, and these invariably lead to much better bounds on mixing time
than known using conductance, and moreover Coupling based proofs
usually turn out to be much simpler.  These applications motivate the
question of whether Coupling indeed can be made to work whenever the
chain is rapidly mixing. This question was answered in the negative in
very recent work by Kumar and Ramesh~\cite{KR99}, who showed that no
Coupling strategy can prove the rapid mixing of the famous
Jerrum-Sinclair chain for sampling perfect and near-perfect matchings
(the chain is known to be rapidly mixing via a canonical paths
argument).

\end{abstract}

\newpage

\section{Introduction}

Suppose $\Omega$ is a large finite set of combinatorial structures
(for example the set of feasible solutions to a combinatorial
optimization problem), and let $\pi$ be a probability distribution on
$\Omega$. The general ``sampling'' problem is then to pick an element
of $\Omega$ at random according to the distribution $\pi$. The \mcmc\
method, which is the subject of our study here, provides an elegant
technique to efficiently solve this general computational task in a
wide variety of contexts.

Sampling problems are inherently interesting, and in addition turn out
to have many computational applications, the most notable ones being:
\begin{itemize}
\item {\em Approximate counting:} Here we want to estimate the size of
$\Omega$ to a very good accuracy. It is well known~\cite{JVV} that,
provided a certain technical condition known as {\em
self-reducibility} is met, almost uniform sampling (that is sampling
from a distribution that is {\em statistically close} to the uniform
distribution) is possible in polynomial time if and only if
approximate counting is. This has been one of the main motivations, at
least from the computer science point of view, behind the rapid
progress that has been made in this area. In particular, for a host of
counting problems including several very hard $\#P$-complete problems,
the \mcmc\ method is the only known approach to approximate the number
of feasible solutions.
\item {\em Statistical physics:} Here the space $\Omega$ represents
possible configurations of statistical mechanical system, and $\pi$ is
a ``natural'' probability distribution on $\Omega$, in which the
probability of a configuration is related to its energy. The task is
to sample configurations according to $\pi$, in order to examine
properties of a ``typical'' physical configuration.
\end{itemize}

In this report, we focus only on the sampling problem and omit the
connections to counting since these involve by now standard
reductions. The \mcmc\ method has been a great success story in
solving sampling problems. It solves the sampling problem by the
following approach. An underlying ``Markov chain'' $\M$ on the state
space $\Omega$ is specified through a stochastic {\em transition
probability matrix} of dimension $|\Omega| \times |\Omega|$ whose
$(x,y)^{\rm th}$ entry specifies the probability $P(x,y)$ that the
chain moves from state $x$ to state $y$ in a single step (we assume
states of $\M$ are labeled by elements of $\Omega$). Starting at any
state $x_0$, there is a natural {\em random walk}
$X_0=x_0,X_1,X_2,\dots$ defined on $\M$ such that $\pr[X_{t+1} |
X_0,\dots,X_t ] = \pr[X_{t+1} | X_t]$ where the latter conditional
probability is specified by the matrix $P$, i.e., $\pr[X_{t+1} = y |
X_t = x] = P(x,y)$. In other words we start at state $X_0$ and at each
time step $t$, we make a move to a next state $X_{t+1}$ by moving to a
random state from the current state $X_t$ according to the transition
probabilities of the chain. Note the crucial ``forgetting property''
of Markov chains: the state at time $t+1$ depends probabilistically on
the state at time $t$, but not on the state at any other time.

To sample according to a distribution $\pi$, the Markov chain $\M$ is
defined in such a way that it is {\em ergodic}, i.e., has a (unique)
stationary distribution $\eta$ on $\Omega$ such that $\pr[X_t=y |
X_0=x] \rightarrow \eta(y)$ as $t \rightarrow \infty$, for all pairs
of states $x,y \in \Omega$, and moreover the transition probabilities
are set up so that $\eta = \pi$. Now we may sample from $\Omega$
according to $\pi$ as follows: starting from an arbitrary state in
$\Omega$, take a random walk on the Markov chain (which we will
loosely refer to as ``simulating the Markov chain'' in the sequel) for
some number, $T$, of steps, and then output the final state. The
ergodicity of $\M$ implies that, by taking $T$ large enough, we can
ensure that the output state is arbitrarily close to the desired
distribution $\pi$.

One of the most appealing things about this method is its simplicity
-- in fact in most applications it is not hard to construct a Markov
chain having the above properties. The crux of the method, which is
also its sticking point, is to obtain good upper bounds on the {\em
mixing time} of the chain, i.e., the number of simulation steps
$T$ necessary before the Markov chain is close to its stationary
distribution. This is critical as this forms the crucial factor in the
running time of any sampling algorithm that uses the chain. Since our
aim is to sample from a set $\Omega$ which is very large, we would
like $T$ to be much smaller than the size of $\Omega$, say at most a
polynomial in the logarithm of $|\Omega|$. We shall refer to such
chains as {\em rapidly mixing}. Over the years several deep and novel
analytic tools have been developed and refined to bound mixing times
of Markov chains.  It is the goal of this report to survey the known
techniques for proving rapid mixing, to present representative
examples of their use, and to compare and contrast their scope, their
relative strengths and limitations, and their applicability to various
contexts.

\medskip \noindent {\bf Organization.} We begin in the next section by
reviewing the relevant definitions and properties of Markov chains,
and by giving a precise characterization of when a Markov
chain mixes rapidly in terms of its {\em spectral} properties. In
Section~\ref{sec:3} we discuss the notion of conductance and its
relation to the spectral gap of the chain. Section~\ref{sec:can-paths}
discusses the canonical paths approach and some of its generalizations
that yield bounds on the conductance and the spectral gap, and also
proves that for a large class of chains a small mixing time implies
the existence of some collection of good canonical paths. We then
present an illustrative application of this technique to the problem
of sampling $0$-$1$ knapsack solutions in
Section~\ref{sec:knapsack-MS}. Section~\ref{sec:coupling} discusses
Coupling which is an entirely different approach to bounding the
mixing time, gives an illustrative example of Coupling in action, and
also discusses Path Coupling, which is a useful design tool in
constructing Couplings. Several elegant applications of Path Coupling
are presented in Section~\ref{sec:path-coupling-applns}. In
Section~\ref{sec:kumar-ramesh} we discuss the recent result of
\cite{KR99} which proves that Coupling is in fact weaker than
conductance, in that there are chains with large conductance which
cannot be shown to be rapidly mixing by {\em any} Coupling
strategy. Finally, we conclude with a few remarks and open questions
in Section~\ref{sec:conclusions}.

\medskip \noindent {\bf Acknowledgments.} This survey was written as
part of the author's Area Examination at MIT, the goal of which was to
survey the papers by Bubley and Dyer~\cite{BD97}, Anil Kumar and
Ramesh~\cite{KR99}, and Morris and Sinclair~\cite{MS99}. This survey
(specifically Sections \ref{sec:knapsack-MS},
\ref{subsec:path-coupling}, \ref{subsec:pc-coloring} and
\ref{sec:kumar-ramesh}) uses liberal portions of the contents of these
papers. This work was also influenced greatly by the reading of the
survey by Jerrum~\cite{jerrum-survey}, and the paper by
Sinclair~\cite{sinclair92}, among several other papers. I would like
to thank Kumar and Ramesh for sending me a copy of the most recent
version of their paper~\cite{KR99}.

\section{Preliminaries on Markov Chains}

A Markov chain on state space $\Omega$ is completely specified by the
{\em transition matrix} $P$ whose entry $P(x,y)$ represents the
probability that the chain moves from state $x$ to state $y$ is a
single transition; i.e., $P(x,y) = \pr[X_{t+1}=y | X_t=x]$ for all $t
\ge 0$. Thus in order to study and analyze the properties of the
Markov chain, it suffices to investigate the properties of this matrix
$P$.

\subsection{Basic definitions}

Starting from an initial distribution $\mu^{(0)}$, the distribution of
the chain after $t$ steps $\mu^{(t)}$ is clearly given by $\mu^{(t)}
= \mu^{(0)} P^n$ (here we view the distributions as row vectors in
$\real^{\Omega}$). Thus, when using a Markov chain to randomly sample
from its state space, we must study the evolution of $\mu^{(t)}$ as
$t$ increases, and we would like $\mu^{(t)}$ to (quickly) approach a limiting
{\em stationary} distribution, say $\pi$; it is not surprising that
$\pi$ must be fixed under steps of the chain.

\begin{defn}
A row vector $\pi \in \real^{\Omega}$ is a {\em stationary
distribution} for a Markov chain $\M$ with transition matrix $P$ if
(a) $\pi(x) \ge 0$ for all $x \in \Omega$, (b) $\sum_{x\in \Omega}
\pi(x) =1$, and (c) $\pi = \pi P$.
\end{defn}

\begin{defn}
A Markov chain $\M$ is said to be {\em ergodic} if it has a stationary
distribution.

\end{defn}

\noindent Clearly, we would like (and need) all Markov chains we use
for sampling to be ergodic, so next we turn to conditions on the chain
which will ensure ergodicity.

\begin{defn}
A Markov chain $\M$ (with transition matrix $P$) is said to be {\em
irreducible} if for all $x,y \in \Omega$, there is an $m$ such that
$P^m(x,y) > 0$, i.e $y$ is eventually reachable from $x$ with non-zero
probability. 
\end{defn}

Irreducibility guarantees that the underlying chain is connected, so
that starting at any state it is possible to reach all the other
states. It is clearly desirable (and necessary) to impose this
requirement when using a Markov chain to sample from a set
$\Omega$. We next impose another condition on the chains we will
study, namely {\em aperiodicity}; this is merely a technical condition
imposed to simplify analysis, and does not cause any loss of
generality as we can turn any (periodic) chain into an aperiodic one
by simply adding loop probabilities of $1/2$ at each state, and this
clearly does not affect the stationary distribution.

\begin{defn}
A chain $\M$ over state space $\Omega$ is {\em aperiodic} iff for all
$x \in \Omega$,
\[ {\rm gcd}\{m : P^m(x,x) > 0 \} = 1 . \]
\end{defn}

A central theorem in the classical theory of stochastic process is the
following:

\begin{theorem}
\label{thm:ergodic}
Any finite, irreducible, aperiodic Markov chain is ergodic.
\end{theorem}

\begin{defn}
Suppose $\M$ (defined over state space $\Omega$) has a stationary
distribution $\pi$. $\M$ is said to be {\em reversible} (with respect
to $\pi$) iff 
\begin{equation}
\label{eq:detailed-balance}
\pi(x) P(x,y) = \pi(y) P(y,x) \mbox{ for all } x,y \in
\Omega \ .
\end{equation}
\end{defn}

The conditions of (\ref{eq:detailed-balance}) are known as {\em
detailed balance equations}. The condition of reversibility does cause
some loss of generality, but the ease of analysis gained by making
this requirement more than compensates the sacrifice made. Moreover,
reversible chains will be general enough for our applications, and for
the rest of the section we focus attention solely on finite,
irreducible, aperiodic and reversible Markov chains.

The detailed balance conditions also permit an easy proof that a
certain distribution is indeed the stationary distribution of an
ergodic Markov chain, as is formalized below.

\begin{lemma}
\label{lem:reversible}
For a Markov chain $\M$ defined on state space $\Omega$, if there
exists a probability distribution $\pi$ on $\Omega$ that satisfies the
conditions (\ref{eq:detailed-balance}), then $\pi$ is a stationary
distribution of $\M$ and $\M$ is reversible with respect to $\pi$.
\end{lemma}
\proof We easily verify that $\pi P = \pi$. Indeed,
\[ (\pi P)(x)  = \sum_y \pi(y) P(y,x) = \sum_y \pi(x) P(x,y) = \pi(x)
\sum_y P(x,y) = \pi(x) \ . \qquad \Box \]

Note that in the definition of ergodicity we did not require the stationary
distribution to be unique, but the conditions of
Lemma~\ref{lem:reversible} together with irreducibility, are
sufficient to guarantee that $\pi$ is in fact the unique stationary
distribution.

\subsection{Spectral theory of reversible Markov chains}

Since a stationary distribution of a Markov chain is simply a left
eigenvector of its transition matrix $P$, it is natural that in order
to study the rate of convergence of the chain to its stationary
distribution, we should try to investigate the spectral properties of
$P$. The reversibility constraint implies that one can view $P$ as a
self-adjoint operator on a suitable inner product space and this
permits us to use the well-understood spectral theory of self-adjoint
operators. This approach was first undertaken in \cite{diaconis} (also
see \cite{vadhan} for a nice exposition).

The relevant inner product space is $L^2(\pi^{-1})$ which is the space
of real-valued functions on $\Omega$, with the following inner
product:\footnote{It is easy to see that the stationary distribution
satisfies $\pi(x) > 0$ for all $x \in \Omega$ whenever the chain is
irreducible, so the inner product is well-defined.}
\begin{equation}
\langle \phi, \psi \rangle = \sum_{x \in \Omega} \frac{\phi(x)
\psi(x)}{\pi(x)}. 
\end{equation}
It is easy to check that the detailed-balance conditions
(\ref{eq:detailed-balance}) imply that $\langle \phi P , \psi \rangle
= \langle \phi, \psi P \rangle$, so that $P$ is a self-adjoint
operator on $L^2(\pi^{-1})$. Now, by standard linear algebra, it is
well known that such a $P$ has $N = |\Omega|$ real eigenvalues $1 =
\lambda_0 > \lambda_1 \ge \lambda_2 \ge \cdots \ge \lambda_{N-1} \ge
-1$; the chain defined by $P$ is ergodic iff $\lambda_{N-1} >
-1$. Also, the space $L^2(\pi^{-1})$ has an orthonormal basis
comprising of eigenvectors $\pi = v_0,v_1,v_2,\dots,v_{N-1}$ of $P$
corresponding to the eigenvalues
$\lambda_0,\lambda_1,\dots,\lambda_{N-1}$.

Now, our initial distribution on $\Omega$ can be written as $\mu^{(0)}
= c_0 \pi + c_1 v_1 + \cdots + c_{N-1} v_{N-1}$ where $c_i = \langle
\mu^{(0)},v_i \rangle$ (so in particular $c_0 = \sum_x
\frac{\mu^{(0)}(x) \pi(x)}{\pi(x)} = 1$). The distribution after $t$
steps is then given by
\begin{equation}
\label{eq:t-step-dist}
\mu^{(t)} = \mu^{(0)} P^t = \pi + c_1 \lambda_1^t v_1 + \cdots +
c_{N-1} \lambda_{N-1}^t v_{N-1}.
\end{equation}

From the above, it is clear that the chain is ergodic whenever
$\lambda_{N-1} > -1$, as then all eigenvalues $\lambda_i$, $1 \le i
\le N-1$, have absolute value less than $1$, and as $t \rightarrow
\infty$, terms corresponding to them will become insignificant, and
$\mu^{(t)} \rightarrow \pi$. For an ergodic chain,
Equation~(\ref{eq:t-step-dist}) also clearly demonstrates that the
rate of convergence to $\pi$ is governed by the second-largest
eigenvalue in absolute value, $\lambda_{\rm max} = \max
\{\lambda_1,|\lambda_{N-1}| \}$. We now make this statement
precise. For $x \in \Omega$, denote by $P^t(x,\cdot)$ the distribution
of the state of the Markov chain at time $t$, when the chain starts at
time $t=0$ in state $x$.
\begin{defn}
The {\em variation distance} at time $t$ with initial state $x$ is
defined as the statistical difference between distributions
$P^t(x,\cdot)$ and $\pi(\cdot)$, i.e
\[ \Delta_x(t) = \frac{1}{2} \sum_{y \in \Omega} | P^t(x,y) - \pi(y) |
. \]
\end{defn}

\noindent We will measure the rate of convergence using the function
$\tau_x$, which quantifies the {\em mixing time}, and which is defined
for $\eps > 0$ by
\begin{equation}
\label{eq:mixing-time}
\tau_x(\eps) = \min \{ t : \Delta_x(t') \le \eps \mbox{ for all } t'
\ge t \} \ .
\end{equation}

\noindent (It is easy to see that if $\Delta_x(t) \le \eps$ then
$\Delta_x(t') \le \eps$ for all $t' \ge t$ as well.) With this
notation, we will say a Markov chain is {\em rapidly mixing} if
$\tau_x(\eps)$ is $O({\rm poly}(\log (N/\eps)))$ (in applications the
number of states $N$ will be exponential in the problem size $n$, so
this amounts to saying that we need to simulate the chain only for
${\rm poly}(n)$ steps in order to get a ``good'' sample from
$\Omega$).  The following makes precise our intuition that a large
value of the {\em spectral gap} $(1-\lambda_{\rm max})$ exactly {\em
captures} the rapid convergence to stationarity. A proof can be found
in \cite{diaconis,aldous}.

\begin{prop}
\label{prop:spectrum}
The quantity $\tau_x(\eps)$ satisfies
\begin{itemize}
\item[(i)] $\tau_x(\eps) \le (1-\lambda_{\rm max})^{-1} \Big( \ln
\pi(x)^{-1} + \ln \eps^{-1} \Big)$.
\item[(ii)] $\max_{x \in \Omega} \tau_x(\eps) \ge \frac{1}{2}
\lambda_{\rm max} (1-\lambda_{\rm max})^{-1} \ln(2\eps)^{-1}$. 
\end{itemize}
\end{prop}

In light of the above Proposition, if we want rapid convergence to the
stationary distribution irrespective of the starting state (which is
desirable for our applications in sampling where we would like to
start at some arbitrary state), a large gap $(1-\lambda_{\rm max})$ is
both a necessary and sufficient condition. Moreover, in practice the
smallest eigenvalue $\lambda_{N-1}$ is unimportant: a crude approach
is to add a holding probability of $1/2$ to every state, i.e., replace
$P$ by $\frac{1}{2}(I + P)$, where $I$ is the $N \times N$ identity
matrix. This ensures that all eigenvalues are positive while
decreasing the spectral gap $(1-\lambda_1)$ only by a factor of
$2$. The upshot is that in order to study mixing times of Markov
chains, one needs to focus attention on the second-largest eigenvalue
$\lambda_1$, and bound it away from $1$. 

\subsection{Characterizations of second-largest eigenvalue}
We now present the known characterizations of the second largest
eigenvalue $\lambda_1$ of self-adjoint matrices, which will be useful
in obtaining good bounds on the spectral gap $(1-\lambda_1)$.

\begin{lemma}[Rayleigh-Ritz]
\label{lem:rayleigh-ritz}
Let $P$ be a self-adjoint operator on a finite-dimensional inner
product space with inner product $\langle \cdot, \cdot
\rangle$. Suppose the eigenvalues of $P$ are $\lambda_0 \ge \lambda_1
\ge \cdots \ge \lambda_{m}$ and $v_0$ is an eigenvector of eigenvalue
$\lambda_0$. Then
\begin{equation}
\label{eq:rayleigh-ritz}
\lambda_1 = \sup_{x \perp v_0} \frac{\langle x,xP \rangle}{\langle x,x
\rangle} \ .
\end{equation}
\end{lemma}
\proof Let $v_0,v_1,\dots,v_m$ be an orthonormal basis of eigenvectors
corresponding to the eigenvalues $\lambda_0,\dots,\lambda_m$
respectively. Since $x \perp v_0$, we can write $x$ as $x = c_1 v_1 +
\cdots + c_m v_m$, so that
\[ \langle x,xP \rangle = \sum_{i=1}^m \lambda_i c_i^2 \le \lambda_1
\sum_{i=1}^m c_i^2 = \lambda_1 \langle x,x \rangle . \]
When $x = v_1$, equality is achieved, and hence the result
follows. {\hfill $\Box$}

\medskip \noindent We next present another characterization which at
first glance seems a bit unwieldy, but it turns out to be quite useful
in that very natural geometrical arguments about a Markov chain can
yield upper bounds on $\lambda_1$ via this
characterization~\cite{diaconis}.

\begin{lemma}[Variational characterization]
\label{lem:variational}
Let $P$ be a self-adjoint operator on a finite-dimensional inner
product space $L^2(\pi^{-1})$, and for $x,y \in \Omega$, let $Q(x,y) =
\pi(x) P(x,y) = Q(y,x)$. Then, the second-largest eigenvalue of $P$
satisfies:
\begin{equation}
\label{eq:variational}
1 - \lambda_1 = \inf_{\psi} \frac{\sum_{x,y \in \Omega} (\psi(x) -
\psi(y))^2 Q(x,y)}{\sum_{x,y \in \Omega} (\psi(x) -
\psi(y))^2 \pi(x) \pi(y)} \ .
\end{equation}
\end{lemma}

\section{Two broad approaches to proving Rapid Mixing}
\label{sec:3}
We saw in the last section that establishing rapid mixing for a Markov
chain amounts to bounding the second largest eigenvalue $\lambda_1$ of
the transition matrix $P$ away from $1$ by a ${\rm poly}(\log N)^{-1}$
amount. The spectrum of the chain is very hard to analyze directly, so
we either need tools to analyze the spectral gap (using the
characterizations presented in the previous section), or somehow
analyze the chain directly without resorting to spectrum.

\subsection{Coupling}
One simple and elegant approach to bound mixing times without
explicitly bounding the spectral gap is {\em Coupling}. A
``coupling'' argument is in fact the classical approach to bound
mixing times of Markov chains. Coupling was first used by
Aldous~\cite{aldous1} to show rapid mixing, and has since found
several applications in proving rapid mixing of a variety of
chains. We will define Coupling formally and discuss some of its
applications in detail in later Sections, but at a very high level the
idea behind Coupling is the following. One sets up two stochastic
processes ${\cal X}=(X_t)$ and ${\cal Y}=(Y_t)$ on the state space
$\Omega$ both of which individually are faithful copies of the Markov
chain $\M$ (whose mixing time we wish to bound). However, their joint
evolution is set up in a way that encourages $(X_t)$ and $(Y_t)$ to
{\em coalesce} rapidly, so that $X_t = Y_t$ for all sufficiently large
$t$. The relevance to rapid mixing is obvious from the Coupling
Lemma~\cite{aldous1,jerrum-survey} which states that the probability
that the coupling time exceeds some value $t$ for a certain
distribution $\pi'$ for $X_0$ is an upper bound on the
variation distance between the stationary distribution $\pi$ of $\M$
and the distribution of the chain at time $t$ starting from
distribution $\pi'$. Note that we did not explicitly deal with the
spectrum of the chain, and this is one advantage of this approach.
We will come back to a detailed discussion of Coupling in Sections
\ref{sec:coupling} through \ref{sec:kumar-ramesh}.

\subsection{Conductance}
Let us now look at approaches aimed at establishing rapid mixing via
directly bounding the spectral gap. These use geometric properties of
the chain and the characterizations of $\lambda_1$ given by Equations
(\ref{eq:rayleigh-ritz}) and (\ref{eq:variational}) to prove a lower
bound on the spectral gap $(1-\lambda_1)$. The relevant geometric
parameter is the {\em conductance} of the chain which is defined
below.

\begin{defn}
\label{def:conductance}
The {\em conductance} of $\M$ is defined by
\begin{equation}
\label{eq:conductance}
\Phi = \Phi(\M) \eqdef \min_{{S \subset \Omega} \atop {0 < \pi(S) \le
1/2}} \frac{Q(S,\bar{S})}{\pi(S)},
\end{equation}
where $Q(x,y) = \pi(x) P(x,y) = \pi(y) P(y,x)$, $\pi(S)$ is the
probability density of $S$ under the stationary distribution $\pi$ of
$\M$, and $Q(S,\bar{S})$ is the sum of $Q(x,y)$ over all $(x,y) \in
S \times (\Omega - S)$.
\end{defn}

The conductance may be viewed as a weighted version of {\em edge
expansion} of the graph underlying the chain $\M$. For a fixed $S$,
the quotient in Equation (\ref{eq:conductance}) is just the
conditional probability that the chain in equilibrium escapes from the
subset $S$ of the state space in one step, given that it is initially
in $S$. Thus $\Phi$ measures the ability of $\M$ to escape from any
small region of the state space, and hence to make rapid progress to
the stationary distribution. It is not therefore very surprising that
the conductance $\Phi$ would govern the rapid mixing
properties of the chain, which in turn is related to the
second-largest eigenvalue $\lambda_1$ (by
Proposition~\ref{prop:spectrum}). This is made precise in the
following result from \cite{sinclair-thesis,SJ89}; related results
appear in \cite{alon86,mihail89,mohar}. Note that the result proves
that the conductance captures mixing rate up to square factors, and
thus obtaining a good lower bound on $\Phi$ is {\em equivalent} to
proving rapid mixing.

\begin{theorem}
The second eigenvalue of a reversible chain satisfies
\begin{equation}
\label{eq:lambda-phi}
1 - 2\Phi \le \lambda_1 \le 1 -\frac{\Phi^2}{2} \ .
\end{equation}
\end{theorem}
\proof We only prove the inequality $(1 - \lambda_1) \le 2\Phi$ which
shows, together with Proposition~\ref{prop:spectrum}, implies that a large
conductance (of the order of $1/{\rm poly}(n)$ where $n$ is the
problem size) is necessary for rapid mixing. Our proof follows the
elegant approach of Alon~\cite{alon86} who proved a similar result for
expansion of unweighted graphs. A proof of the other direction:
$(1-\lambda_1) \ge \frac{\Phi^2}{2}$, can be found in
\cite{SJ89,mihail89}.

In order to prove $\lambda_1 \ge (1-2\Phi)$, we use the
characterization of Equation (\ref{eq:rayleigh-ritz}). The largest
eigenvalue of $P$ equals $1$ and has $\pi$ as its eigenvector. Define
a vector $f \in \real^{\Omega}$ (specified as a real-valued function
on $\Omega$) as follows:
\[ f(x) = \left\{ \begin{array}{ll}
\pi(x)\pi(\bar{S}) & \mbox{ if $x \in S$} \\
-\pi(x)\pi(S) & \mbox{ if $x \notin S$}
\end{array} \right. \]
Note that $\langle f, \pi \rangle = \sum_x \frac{f(x) \pi(x)}{\pi(x)}
= \sum_x f(x) = 0$, hence by Equation (\ref{eq:rayleigh-ritz}), we have 
\begin{equation}
\label{eq:tempo-0}
\frac{\dprod{f}{fP}}{\dprod{f}{f}} \le \lambda_1 \ .
\end{equation}
Define $g(x) \eqdef \frac{f(x)}{\pi(x)}$. Now
\begin{eqnarray}
\dprod{f}{f} & = & \sum_x \frac{f^2(x)}{\pi(x)} = \sum_x g^2(x) \pi(x)
	\nonumber \\ & = & \sum_{x \in S} \pi(\bar{S})^2 \pi(x) +
	\sum_{x \in \bar{S}} (-\pi(S))^2 \pi(x) = \pi(S)\pi(\bar{S})
	\label{eq:tempo-1}
\end{eqnarray}
\begin{eqnarray}
\dprod{f}{fP} & = & \sum_x \frac{f(x) \sum_y f(y) P(y,x)}{\pi(x)}
 =  \sum_{x,y} g(x) g(y) Q(x,y) \nonumber \\
& = & \sum_x g^2(x) \sum_y Q(x,y) +
\sum_{x,y} g(x) (g(y) - g(x)) Q(x,y) \nonumber \\
& = & \sum_x g^2(x) \pi(x) + \Big(\sum_{x \in S} - \pi(\bar{S}) (\pi(S)
+\pi(\bar{S})) \sum_{y \in \bar{S}} Q(x,y)\Big) \nonumber \\
& & \qquad\qquad\quad\quad + \Big( \sum_{x \in \bar{S}}
-\pi(S) (\pi(\bar{S}) + \pi(S)) \sum_{y \in S} Q(x,y)\Big) \nonumber \\
& = & \dprod{f}{f} - Q(S,\bar{S})  =  \pi(S)\pi(\bar{S}) - Q(S,\bar{S}) . \label{eq:tempo-2}  
\end{eqnarray}
From (\ref{eq:tempo-0}), (\ref{eq:tempo-1}) and (\ref{eq:tempo-2}), we
get that for any set $S$,
\[ \frac{Q(S,\bar{S})}{\pi(S)\pi(\bar{S})} \ge 1-\lambda_1 . \]
Since $\pi(\bar{S}) \ge 1/2$, this implies $\Phi \ge
\frac{1-\lambda_1}{2}$, as desired. {\hfill $\Box$}

\begin{cor}
\label{cor:conductance}
Let $\M$ be a finite, reversible, ergodic Markov chain with loop
probabilities $P(x,x) \ge 1/2$ for all states $x$, and let $\Phi$ be
the conductance of $\M$. Then the mixing time of $\M$ satisfies
$\tau_x(\eps) \le 2\Phi^{-2} (\ln \pi(x)^{-1} + \ln \eps^{-1})$.
\end{cor}

\medskip A direct analysis of the conductance is sometimes possible by
exploiting an underlying geometric interpretation of $\M$, in which
states of $\M$ are identified with certain polytopes, and transitions
with their common facets. A lower bound on conductance then follows
from an appropriate ``isoperimetric inequality'' of the graph under
consideration. This has been fruitful in a few applications, for
example the estimation of the volume of a convex body by Dyer, Frieze
and Kannan~\cite{DFK91}, and a Markov chain over linear extensions of
a partial order by Karzanov and Khachiyan~\cite{KK90}.  A more recent
example where the conductance is tackled ``directly'' is the work of
Dyer, Frieze and Jerrum~\cite{DFJ-indset} who prove an upper bound on
$\Phi$ to show that certain classes of Markov chains for sampling
independent sets in sparse graphs do not mix rapidly.  The conductance
is still not very amenable to computation in general, and we need
further tools that can be used to deduce good lower bounds on the
conductance. It is this task to which we turn next.

\section{Rapid mixing via canonical paths}
\label{sec:can-paths}
We saw in the last section that in order to prove rapid mixing of a
Markov chain, all we need is a good lower bound on the conductance (and
hence the spectral gap) of the chain. In this section, we explore a
useful piece of technology developed in
\cite{JS-permanent,sinclair-thesis,sinclair92} to prove such a lower bound. The
basic idea behind the method is to try and associate {\em
canonical paths} between every pair of states, in such a way that no
transition of the chain is used by too many paths. Intuitively, if
such a set of paths exists, this means that the chain has no severe
bottlenecks which could impede mixing. We now turn to formalizing this
intuition.

\subsection{Bounding Conductance using Canonical paths}

We first formalize some terminology and notation. Let $\M$ be an
ergodic Markov chain on a finite set $\Omega$. We define the weighted
directed graph $G(\M)$ with vertex set $\Omega$ and with an edge $e$
between an ordered pair $(x,y)$ of weight $Q(e) = Q(x,y) = \pi(x)
P(x,y)$ whenever $P(x,y) > 0$. We call this the underlying graph of
$\M$.

A set of {\em canonical paths} for $\M$ is a set $\Gamma$ of simple
paths $\{\gamma_{xy}\}$ in the graph $G(\M)$, one between each ordered
pair $(x,y)$ of distinct vertices. In order to bound the conductance,
we would like to have a set of canonical paths that do not overload
any transition of the Markov chain. To measure this ``overloading'',
we define the {\em path congestion}
parameter~\cite{JS-permanent,sinclair-thesis} for a set of canonical
paths $\Gamma$ as:
\begin{equation}
\label{eq:path-congestion}
\rho(\Gamma) = \max_{e \in G(\M)} \frac{1}{Q(e)} \sum_{\gamma_{xy} \ni
e} \pi(x) \pi(y),
\end{equation}
where the maximum is over all oriented edges $e$ in $G(\M)$, and $Q(e)
= Q(x,y)$ if $e=(x,y)$. Think of the Markov chain as a flow network in
which $\pi(x)\pi(y)$ units of flow travel from $x$ to $y$ along
$\gamma_{xy}$, and $Q(e)$, which equals the probability that the Markov
chain in the stationary distribution will use the transition $e$ in a
single step, serves as the capacity of $e$. The quantity $\rho(\Gamma)$
measures the maximum overloading of any edge relative to its
capacity. The path congestion $\rho = \rho(\M)$ of the chain $\M$ is
defined as the minimum congestion achievable by any set of canonical
paths, i.e.,
\begin{equation}
\label{eq:pc}
\rho = \inf_{\Gamma} \rho(\Gamma).
\end{equation}

The following simple result confirms our intuition that a set of paths
with low congestion implies a large value of conductance.

\begin{lemma}
\label{lem:pc-implies-conductance}
For any reversible Markov chain and any set of canonical paths
$\Gamma$, we have 
\[ \Phi \ge \frac{1}{2\rho(\Gamma)} \ . \]
\end{lemma}
\proof Pick $S \subset \Omega$ with $0 < \pi(S) \le 1/2$ such that
$\Phi = \frac{Q(S,\bar{S})}{\pi(S)}$. For any choice of paths
$\Gamma$, the total flow from $S$ to $\bar{S}$ is
$\pi(S)\pi(\bar{S})$, and all this must flow across the cut
$[S:\bar{S}]$, which has capacity 
$Q(S,\bar{S})$. Hence there must exist an edge $e$ in the cut
$[S:\bar{S}]$ such that
\[ \frac{1}{Q(e)} \sum_{\gamma_{xy} \ni e} \pi(x)\pi(y) \ge
\frac{\pi(S)\pi(\bar{S})}{Q(S,\bar{S})} \ge
\frac{\pi(S)}{2Q(S,\bar{S})} = \frac{1}{2\Phi} \ . \qquad\quad \Box \]

\begin{cor}
\label{cor:pc-spectrum}
For any reversible Markov chain, and any choice of canonical paths
$\Gamma$, the second-largest eigenvalue $\lambda_1$ satisfies 
\begin{equation}
\label{eq:pc-spectrum}
\lambda_1 \le 1 - \frac{1}{8\rho^2(\Gamma)} \ . 
\end{equation}
\end{cor}

\subsection{Relating Spectrum to congestion directly}

Since the relation between $\rho$ and $(1-\lambda_1)$ above proceeded
by appealing to the conductance, the bound of
Corollary~\ref{cor:pc-spectrum} is potentially rather weak because of
the appearance of the square. So we now pursue a direct approach to
bound $\lambda_1$ based on the existence of ``good'' canonical
paths. This was first achieved by Diaconis and Strook~\cite{diaconis},
but we follow a treatment by Sinclair~\cite{sinclair92} as it gives
the best bounds for the examples considered later.

In order to state the new bound on $\lambda_1$, we modify the measure
$\rho(\Gamma)$ to take into account the lengths of the paths as
well. For a set $\Gamma = \{\gamma_{xy}\}$ of canonical paths, the key
quantity is now
\begin{equation}
\label{eq:lpc}
\bar{\rho}(\Gamma) = \max_{e} \frac{1}{Q(e)} \sum_{\gamma_{xy} \ni e} \pi(x)\pi(y)|\gamma_{xy}|,
\end{equation}
where $|\gamma_{xy}|$ stands for the length of the path
$\gamma_{xy}$. The parameter $\bar{\rho}$ is defined analogously to
Equation~(\ref{eq:pc}) by minimizing over the choice of $\Gamma$.

Note that it is reasonable to allow general length functions $l(e)$ on
the edges $e$, compute $|\gamma_{xy}|$ in terms of this length
function, and thus obtain a quantity similar to $\bar{\rho}(\Gamma)$
above. In fact, Diaconis and Strook use the length function $l(e) =
1/Q(e)$, and Kahale~\cite{kahale} considers good length functions that
will lead to the best bounds for specific chains. We will be content
with the unit length function for our purposes.

Intuitively, the existence of short paths which do not overload any
edge should imply that the chain mixes very rapidly. Indeed, it turns
out that the variational characterization (\ref{eq:variational}) can
now be used to bound $\lambda_1$ directly in terms of
$\bar{\rho(\Gamma)}$. This is stated in the theorem below; we will not
prove this theorem, but will later prove a more general version of
this result (namely Theorem~\ref{thm:lfc-spectrum}, which also appears in \cite{sinclair92}.

\begin{theorem}[\cite{sinclair92}]
\label{thm:lpc-spectrum}
For any reversible Markov chain, and any choice of canonical paths
$\Gamma$, the second-largest eigenvalue $\lambda_1$ satisfies
\begin{equation}
\label{eq:lpc-spectrum-1}
\lambda_1 \le 1 - \frac{1}{\bar{\rho}(\Gamma)} \ .
\end{equation}
\end{theorem}

A useful way to use the above result is the following version which
bounds the spectral gap in terms of the path congestion $\rho$ and the
length of a longest path used in $\Gamma$. This version of the result
is the most used in bounding mixing times using this approach.

\begin{cor}
\label{cor:lpc-spectrum}
For any reversible Markov chain, and any choice of canonical paths
$\Gamma$, the second-largest eigenvalue $\lambda_1$ satisfies
\begin{equation}
\label{eq:lpc-spectrum-2}
\lambda_1 \le 1 - \frac{1}{\rho(\Gamma) \ell} \ .
\end{equation}
where $\ell =\ell(\Gamma)$ is the length of a longest path in $\Gamma$.
\end{cor}

The above often leads to much sharper bounds on mixing times than
(\ref{eq:pc-spectrum}) because the maximum path length $\ell$ will
usually be significantly lesser than the estimate obtained for $\rho$.

\subsection{Known applications of canonical paths}
The ``canonical paths'' approach has been applied successfully to
analyze a variety of Markov chains including those for sampling
perfect matchings and approximating the
permanent~\cite{JS-permanent,diaconis}, estimating the partition
function of the Ising model~\cite{JS-ising}, sampling bases of
balanced matroids~\cite{FM92}, sampling regular bipartite
graphs~\cite{KTV97}, sampling $0$-$1$ knapsack
solutions~\cite{DFKKPV}, etc. All these papers with the exception of
\cite{FM92} use more or less the same technique to bound the path
congestion that is due to \cite{JS-permanent} -- they use the
state space to somehow ``encode'' the paths that use any given
transition, so that the number of paths through any edge will be
comparable to the number of states of the chain. Feder and
Mihail~\cite{FM92} give a random collection of canonical paths and use
a variant of ``Hall's condition'' (for existence of perfect matchings
in bipartite graph) to show a small expected congestion and maximum
path length for this collection of paths. They also prove a version of Corollary
\ref{cor:lpc-spectrum} which applies with expected path lengths and
congestion instead of worst case values.

\subsection{Path congestion is weaker than Conductance}
\label{sec:example}

The canonical paths technique is very useful, but it is natural to ask
whether, like conductance, it too {\em captures} rapid mixing up to
some polynomial factor (recall that conductance captures mixing time
up to square factors). In other words, does a large conductance or a
large spectral gap $(1-\lambda_1)$ always imply a small value of
$\rho(\Gamma)$ for some choice of canonical paths $\Gamma$?
Unfortunately we give a simple example below to show that the answer
is no --- the same example also appears in \cite{sinclair92}.

\medskip \noindent {\bf Example.} Consider the complete bipartite
graph $K_{2,n-2}$ on vertex set $\{1,2,\dots,n\}$ and edges
$\{(1,i),(2,i) : 3 \le i \le n\}$ where $n$ is even, and define
transition probabilities corresponding to the random walk on this
graph, namely at each step stay where you are with probability $1/2$,
else move to a neighbor chosen uniformly at random. The stationary
distribution $\pi$ of this Markov chain is given by: $\pi(1)=\pi(2) =
1/4$ and $\pi(i) = 1/2(n-2)$ for $i=3,4,\dots,n$, and hence $Q(e) =
1/4(n-2)$ for all edges $e$.  Since $n$ is even it is easy to verify
that the conductance of this chain is $\Phi = 1/2$, and hence using
Equation~(\ref{eq:lambda-phi}) we get $\lambda_1 \le 7/8$. However,
since $\pi(1)\pi(2) = 1/16$ and $Q(e)=1/4(n-2)$ for all edges $e$, the
path connecting states $1$ and $2$ alone implies that the best value
for $\rho(\Gamma)$ or $\bar{\rho}(\Gamma)$ obtainable using canonical
paths is $\Omega(n)$. Hence $\rho$ and $\bar{\rho}$ could in fact be
much larger than the quantity $(1-\lambda_1)^{-1}$ which governs the
mixing time.

\subsection{Resistance: a generalization of path congestion}

In order to alleviate the shortcoming of the canonical paths technique
which was just discussed, we now present a natural generalization of
this approach that will end up capturing mixing times exactly
(and will thus be ``as good as'' conductance). The idea, again due to
Sinclair~\cite{sinclair92}, is to {\em spread the flow} on path
$\gamma_{xy}$ between a pair $(x,y)$ of states among {\em several
paths}. As before, we view $G(\M)$ as a flow network where one unit of
flow has to be routed from $x$ to $y$ for every ordered pair $(x,y)$ of distinct vertices, and each (oriented) edge $e$ has ``capacity''
$Q(e)$. The difference from the canonical paths approach is that, we
now allow the flow between $x$ and $y$ to be {\em split} among
multiple paths, i.e., we are looking for a {\em fractional
multicommodity flow} that minimizes the congestion. Considering the
similarity with the earlier approach, it is natural to suppose that
this new measure will yield similar bounds on the mixing rate. As we
shall see, this will be the case, and in fact this seemingly innocuous
generalization to multiple paths allows us to capture rapid mixing
exactly!

Formally, a {\em flow} in $G(\M)$ is a function $f : {\cal P}
\rightarrow \real^{+}$ which satisfies
\[
\sum_{p \in {\cal P}_{xy}} f(p) = 1 \qquad \mbox{for all $x,y \in X$,
$x \neq y$}, \]
where ${\cal P}_{xy}$ is the set of all simple directed paths from $x$
to $y$ in $G(\M)$ and ${\cal P} = \cup_{x\neq y} {\cal P}_{xy}$. The
quality of a flow is measured by the {\em congestion parameter}
$\res(f)$, defined analogously to Equation~(\ref{eq:path-congestion}) by
\begin{equation}
\label{eq:res-1}
\res(f) \eqdef \max_{e} \frac{1}{Q(e)} \sum_{x,y} \sum_{p \in {\cal P}_{xy}: p
\ni e} \pi(x)\pi(y)f(p) ,
\end{equation}
and one can define {\em elongated congestion} $\bar{\res}(f)$, similar
to Equation~\ref{eq:lpc}, by accounting for the lengths of the paths:
\begin{equation}
\label{eq:res-2}
\bar{\res}(f) \eqdef \max_{e} \frac{1}{Q(e)} \sum_{x,y} \sum_{p \in {\cal P}_{xy}: p
\ni e} \pi(x)\pi(y)f(p)|p| .
\end{equation}
\noindent We have the following results parallel to those of
Lemma~\ref{lem:pc-implies-conductance},
Corollary~\ref{cor:pc-spectrum}, Theorem~\ref{thm:lpc-spectrum} and
Corollary~\ref{cor:lpc-spectrum}.

\begin{lemma}
\label{lem:fc-spectrum}
For any reversible Markov chain and any flow $f$, we have
\[ \Phi \ge \frac{1}{2\res(f)} \quad \mbox{and hence} \quad \lambda_1 \le 1 - \frac{1}{8\res(f)^2}\ . \]
\end{lemma}

\begin{theorem}
\label{thm:lfc-spectrum}
For any reversible Markov chain, and any flow $f$, the second-largest
eigenvalue $\lambda_1$ satisfies
\begin{equation}
\label{eq:lfc-spectrum-1}
\lambda_1 \le 1 - \frac{1}{\bar{\res}(f)} \ .
\end{equation}
\end{theorem}

\begin{cor}
\label{cor:lfc-spectrum}
For any reversible Markov chain, and any flow $f$,
the second-largest eigenvalue $\lambda_1$ satisfies
\begin{equation}
\label{eq:lfc-spectrum-2}
\lambda_1 \le 1 - \frac{1}{\res(f) \ell(f)} \ .
\end{equation}
where $\ell(f)$ is the length of a longest path $p$ with $f(p) > 0$.
\end{cor}

We now provide a proof of Theorem~\ref{thm:lfc-spectrum} as we had
promised before the statement of Theorem~\ref{thm:lpc-spectrum} (note
that the statement of Theorem~\ref{thm:lfc-spectrum} clearly
generalizes that of Theorem~\ref{thm:lpc-spectrum}).

\medskip \noindent {\bf Proof of Theorem~\ref{thm:lfc-spectrum}:} 
We need to prove $(1-\lambda_1) \ge 1/\bar{\res}(f)$ for any flow
$f$. We use Equation (\ref{eq:variational}) to bound
$(1-\lambda_1)$, namely
\begin{equation}
\label{eq:op}
1 - \lambda_1 = \inf_{\psi} \frac{\sum_{x,y \in \Omega} (\psi(x) -
\psi(y))^2 Q(x,y)}{\sum_{x,y \in \Omega} (\psi(x) -
\psi(y))^2 \pi(x) \pi(y)}.
\end{equation}
Now for any $\psi$, and any flow $f$, the denominator in the above
expression can be written as:
\begin{eqnarray*}
\sum_{x,y \in \Omega} (\psi(x) - \psi(y))^2 \pi(x) \pi(y) & = &
\sum_{x,y} \pi(x) \pi(y) (\psi(x) - \psi(y))^2  \sum_{p \in {\cal
P}_{xy}} f(p) \\
& = & \sum_{x,y} \pi(x) \pi(y) \sum_{p \in {\cal P}_{xy}} 
f(p) \Big( \sum_{e \in p} (\psi(e^+) - \psi(e^-)) \Big)^2  \\
& \le & \sum_{x,y} \pi(x) \pi(y)\sum_{p \in {\cal P}_{xy}} 
f(p) |p| \sum_{e \in p} (\psi(e^+) - \psi(e^-))^2
\\ & = & \sum_{e} (\psi(e^+) - \psi(e^-))^2 \sum_{x,y} \sum_{p \in
{\cal P}_{xy}: p \ni e} \pi(x) \pi(y) f(p) |p| \\
& \le &  \sum_{e} (\psi(e^+) - \psi(e^-))^2 Q(e) \bar{\res}(f) \\
& = & \bar{\res}(f) \sum_{x,y} Q(x,y)(\psi(x) - \psi(y))^2 . 
\end{eqnarray*}
(Here $e^-$ and $e^+$ denote the start and end vertices of the
oriented edge $e$, and we have used Cauchy-Schwartz inequality in the
third step above.) The result now follows from (\ref{eq:op}). {\hfill
$\Box$}

\begin{defn}[Resistance]
\label{def:resistance}
The {\em resistance} $\res =
\res(\M)$ of chain $\M$ is defined as the minimum value of $\res(f)$
over all flows $f$, and like the conductance is an invariant of the
chain. Formally,
\begin{equation}
\label{eq:resistance-defn}
\res = \inf_f \res(f) \ . 
\end{equation}
\end{defn}

\subsection{Resistance captures rapid mixing}

By Lemma~\ref{lem:fc-spectrum}, note that $\lambda_1 \le 1 -
\frac{1}{8\res^2}$, so a small resistance leads to rapid mixing.  We
will now see that in fact the converse is true, in other words a small
mixing time implies a small resistance, i.e., the existence of a flow
$f$ with small congestion $\res(f)$. Thus resistance {\em overcomes}
the shortcoming of path congestion (since low path congestion was not
a necessary condition for rapid mixing, as was shown by the example in
Section~\ref{sec:example}).

\begin{theorem}[\cite{sinclair92}]
\label{thm:res-mixing}
Consider an irreducible, reversible, ergodic Markov chain $\M$ over $\Omega$ and let
$\tau = \max_{x \in \Omega} \tau_x(1/4)$. Then the resistance
$\res=\res(\M)$ of $\M$ satisfies $\res \le 16\tau$.
\end{theorem}
\proof We will demonstrate a flow $f$ with $\res(f) \le 16\tau$. Let
$t=2\tau$. The flow between $x$ and $y$ will be routed as follows:
Consider the set ${\cal P}^{(t)}_{xy}$ of all (not necessarily simple)
paths of length $t$ from $x$ to $y$ in $G(\M)$, and for each $p \in
{\cal P}_{xy}^{(t)}$ route $f(p) \propto {\rm prob}(p)$ units of flow
on $p$, where ${\rm prob}(p)$ is the probability that the Markov chain
makes the sequence of transitions defined by $p$ in the first $t$
steps when starting in state $x$. Since $t = 2\tau$, it is easy to see
that for any pair $x,y$, ${\cal P}^{(t)}_{xy} \neq \emptyset$, and in
fact
\begin{equation}
\label{eq:kl-0}
\frac{P^t(x,y)}{\pi(y)} \ge \frac{1}{8} \ . 
\end{equation}
Thus for $p \in {\cal P}^{(t)}_{xy}$, we have $f(p) = {\rm
prob}(p)/(\sum_{q \in {\cal P}^{(t)}_{xy}} {\rm prob}(q)) = {\rm
prob}(p)/P^t(x,y)$. Now let us estimate the $\res(f)$.
\begin{eqnarray*}
\res(f) & = & \max_e \frac{1}{Q(e)} \sum_{x,y} \sum_{p \in {\cal
P}^{(t)}_{xy}: p \ni e} \frac{\pi(x)\pi(y){\rm prob}(p)}{P^t(x,y)} \\
& \le & \max_e \frac{8}{Q(e)} \sum_{x,y} \sum_{{p \in {\cal
P}^{(t)}_{xy}} \atop {p \ni e}} \pi(x){\rm prob}(p) \quad \mbox{(using
(\ref{eq:kl-0}))} \\
& \le &  \max_e \frac{8}{Q(e)} \cdot t Q(e)  =  8t  = 16\tau
\end{eqnarray*}
where we used the fact that the final double summation is simply the
probability that the Markov chain traverses the edge $e$ within $t$
steps when started in the stationary distribution $\pi$ over $\Omega$,
and this probability, by the union bound, is at most $t$ times the
probability that this happens in one step, and is thus at most $t Q(e)$. {\hfill
$\Box$}

\medskip \noindent {\bf Remark A.} It is also possible to prove (see
\cite{sinclair92}), using techniques of the approximate max-flow
min-cut theorem for uniform multicommodity flow~\cite{LR88}, that
$\lambda_1 \ge 1 - O(\frac{\log N}{\res})$. This gives the weaker bound $\tau
= \Omega(\res/\log N)$, but is interesting in its own right.

\medskip \noindent {\bf Remark B.} Note that since we used paths of
length $2\tau$ in the above proof, the flow $f$ also satisfies
$\bar{\res}(f) = O(\tau^2)$. This, together with (\ref{eq:lfc-spectrum-1}), implies
that $\bar{\res} = \inf_f \bar{\res}(f)$ captures rapid mixing as
well. The work of Kahale~\cite{kahale} actually
shows that the bound on $\bar{\res}(f)$, call it $\mu$, obtained by
minimizing over all length functions on the transitions and all flows,
can be computed to arbitrary precision by reduction to a semidefinite
program, and satisfies $\lambda_1 \ge 1 - O(\frac{\log^2 N}{\mu})$.

\subsection{Path congestion almost always captures rapid mixing!}

In the next section, we will see a resistance based
proof (due to \cite{MS99}) of rapid mixing of a natural Markov chain
for sampling $0$-$1$ Knapsack solutions. This problem was open for a
long time, and had defied all attempts to prove rapid mixing based on
canonical paths. In light of the example in Section~\ref{sec:example},
it is natural to ask if this chain (which we now know mixes rapidly)
also cannot have low path congestion, and whether the generalization
to resistance was really necessary. 

In this section, we will show that, for a broad class of Markov chains,
including all the ones we consider in applications here, the path
congestion $\rho$ (defined in Equations (\ref{eq:path-congestion}) and
(\ref{eq:pc})) {\em characterizes rapid mixing} up to polynomial (in
the problem size) factors. We show that if you can achieve low
congestion with multiple paths, i.e., if the chain has low resistance,
then you can also achieve low congestion by
routing all the flow on just a single path. The proof is actually very
simple, and is based on randomized rounding to relate the optimum
congestion of ``fractional'' and ``unsplittable'' flows, but we were
surprised that it does not seem to have been observed or made explicit
in the literature. 

\begin{theorem}
\label{thm:rand-rounding}
Consider an ergodic, reversible Markov chain $\M$ with stationary
distribution $\pi$ on a state space $\Omega$ of size $N$, and let the
resistance of $\M$ be $\res$. Let ${\Lambda} = \max_{x \neq y} \pi(x)
\pi(y)$, and let $Q_{\min} = \min_{e: Q(e) > 0} Q(e)$. Then there
exists a set of canonical paths $\Gamma$ such that
\[ \rho(\Gamma) = O\Big(\res + \log N \frac{{\Lambda}}{Q_{\min}} \Big)  . \] 
\end{theorem}
\proof By the definition of the resistance $\res$, we know that there
exists a flow $f$ which routes $\pi(x)\pi(y)$ units of flow between
every ordered pair $(x,y)$ of distinct states $x \neq y$, such that
every (oriented) edge $e$ has at most $Q(e) \res$ units of flow
passing through it. Hence there is a feasible fractional flow $f$
which routes $f_{xy} = \pi(x)\pi(y)/{\Lambda} \le 1$ units of flow between
$x$ and $y$, and with ``capacity'' on edge $e$ at most $C(e) =
\max\{\frac{Q(e) \res}{\Lambda},1\}$. We can now use a
result of Raghavan and Thompson~\cite{raghavan}, who used randomized
rounding to show the following: There are absolute constants $b_0$
and $b_1$ such that if all edge capacities equal $1$, and all
demands are at most $1$, and there is a fractional flow satisfying all
the demands with congestion on edge $e$ at most $\mu^f(e) \ge 1$, then there is an {\em
unsplittable} flow which satisfies all the demands by routing the
demand for each source-destination pair along a {\em single} path, and
which has congestion at most $b_0 \mu^f(e) + b_1 \log N$ on edge $e$.

Applying this to our situation with $\mu^f(e)=C(e)$, we conclude that
there exists a set $\Gamma$ of canonical paths which can route $f_{xy}$
units of flow from $x$ to $y$ such that at most $b_0 C(e) + b_1 \log
N$ units flow through any edge $e$, or equivalently, it can route
$\pi(x)\pi(y) = {\Lambda} f_{xy}$ units of flow between every pair
$(x,y)$ such that at most $b_0 {\Lambda} C(e) + b_1 {\Lambda} \log N$
units flow through any edge $e$. This implies that
\[ \rho(\Gamma) \le b_0 \max \{ \res, \frac{\Lambda}{Q_{\min}} \}
+b_1\frac{\Lambda}{Q_{\min}} \log N   \ . \]
and the stated result follows. {\hfill $\Box$}

\bigskip \noindent Theorem~\ref{thm:rand-rounding} actually implies
that $\rho = O(\res)$ for a wide variety of Markov chains, and thus
for these chains $\rho$ also characterizes rapid mixing. Indeed, this
will be the case whenever $\log N \frac{{\Lambda}}{Q_{\min}} =
O(1)$, which will normally always be the case unless the stationary
distribution varies widely in the mass it gives to points of the state
space, or there are very small non-zero transition probabilities in
the chain. As an example consider Markov chains with uniform
stationary distribution. Then $\log N \frac{{\Lambda}}{Q_{\min}} =
O(1)$ whenever $P(x,y) = \Omega(\frac{\log N}{N})$ for all $x,y$ such
that $P(x,y) > 0$. For most chains in applications to sampling, we
will have $N = 2^{O(n)}$ where $n$ is the problem size and each
non-zero $P(x,y)$ will be at least $1/{\rm poly}(n)$, hence this
condition will indeed be met.

\section{Sampling $0$-$1$ Knapsack solutions}
\label{sec:knapsack-MS}
We describe an example of random walk on the truncated hypercube which
was only very recently shown to be rapidly mixing using a fractional
multicommodity flow with low congestion~\cite{MS99}, but had resisted
all efforts of proving such a result using canonical paths (with just
one path between every source-destination pair). Our result from the
previous section (Theorem~\ref{thm:rand-rounding}) applies to this
chain; this shows that even though spreading flow across multiple
paths might in principle be not more powerful than sending all the
flow along a single canonical path, it could be still be easier to
deal with in actually designing the flow. (The example from this
section is also covered by the framework of what Feder and
Mihail~\cite{FM92} did, where they prove a version of the small path
congestion implies small mixing time result using expected path
lengths and congestion instead of worst case values.)

\medskip \noindent {\bf The Problem.} We are interested in sampling
from the set $\Omega$ of feasible solutions to the $0$-$1$ knapsack
problem defined by the vector ${\bf a}$ of item sizes and the knapsack
capacity $b$; i.e., for a positive real vector ${\bf a} =
(a_i)_{i=1}^n$ and a real number $b$,
\[\Omega = \Omega_{{\bf a},b} = \{ {\bf x} \in \{0,1\}^n : {\bf a} \cdot {\bf x} =
\sum_{i=1}^n a_i x_i \le b\} \ . \] 
There is a one-one correspondence between vectors ${\bf x} \in \Omega$
and subsets $X$ of items whose aggregated weight does not exceed $b$,
given by $X = \{i : x_i=1\}$. We will write $a(X)$ for the weight of
$X$, i.e., $a(X) = \sum_{i \in X} a_i$.

A particularly simple Markov chain $\M_{K}$ on $\Omega$ has been proposed for
the purposes of sampling uniformly at random from $\Omega$. If the
current state is $X \subseteq \{1,2,\dots,n\}$ then 
\begin{enumerate}
\item With probability $1/2$ stay at $X$ (this holding probability is
to make the chain aperiodic), else
\item Pick an item $i \in \{1,2,\dots,n\}$ uniformly at random. If $i
\in X$ move to $X - \{i\}$; if $i \notin X$ and $a(X \cup \{i\}) \le
b$, move to $X \cup \{i\}$, else stay at $X$.
\end{enumerate}

The chain is aperiodic since $P(X,X) \ge 1/2$ for all states $X$, and
it is irreducible since every pair of states can be connected via the
empty set. Moreover, it is clear that each non-zero transition
probability $P(X,Y)$, $X \neq Y$, equals $P(X,Y) = P(Y,X) =
\frac{1}{2n}$. By Theorem~\ref{thm:ergodic} and
Lemma~\ref{lem:reversible} therefore, $\M_K$ is ergodic with uniform
stationary distribution. Despite all the recent activity in proving
rapid mixing, this simple example was not known to be rapidly mixing until
the work of \cite{MS99}. The best prior known bound on the mixing time,
obtained via the canonical paths technique, was ${\rm exp}(O(\sqrt{n}
(\log n)^{5/2}))$~\cite{DFKKPV}, which beats the trivial bound of
${\rm exp}(O(n))$ but is still exponential. 

We will now sketch the proof of \cite{MS99} that this chain has a
mixing time of $O(n^8)$, and is thus indeed rapidly mixing. The proof
will follow the {\em resistance} approach, i.e., we will find a flow
$f$ that routes one unit of flow between every pair of unequal states,
using multiple paths for each pair to ``spread'' the flow, and then
use Corollary~\ref{cor:lfc-spectrum} to bound the mixing time. Indeed,
if $L(f)$ is the length of the longest flow carrying path, and $C(f)$
is the maximum flow across any (oriented) edge of the chain, then
combining Corollary~\ref{cor:lfc-spectrum} and
Proposition~\ref{prop:spectrum} shows that
\begin{equation}
\label{eq:mixing-time-mk}
\tau_X(\eps) \le 2n
\frac{C(f)}{|\Omega|} L(f) (n + \ln \eps^{-1}) \ . 
\end{equation}

Hence our goal now is to construct a flow $f$ with $L(f) = {\rm
poly}(n)$ and $C(f) = |\Omega| {\rm poly}(n)$. Note that a shortest
path between states $X$ and $Y$ can be viewed as a permutation of the
symmetric difference $X \xor Y$, the set of items that must be added
to or removed from the knapsack in passing from $X$ to $Y$. A natural
approach to defining a good flow seems to be to spread the unit flow
from $X$ to $Y$ evenly among all permutations of $X \xor Y$. The problem with
this approach, however, is that many of these permutations will tend
to violate the knapsack constraint, as too many items will have been
added at some intermediate point; i.e., the permutation is
unbalanced. The way to circumvent this problem is to define a family
of permutations, which are all ``balanced'' and also ``sufficiently
random'', and spread the flow evenly among them. Proving the existence
of such permutations, called {\em balanced almost uniform
permutations} in \cite{MS99}, forms the main technical component of this proof. 

We will now define the notion of \baup\ formally, and state the
Theorems from \cite{MS99} guaranteeing their existence. (We will not
prove these theorems as they are quite technical and doing so will
take us too far away from our main theme of focusing on Markov chain
techniques.) We will, however, show how to construct a good flow $f$
for our purposes given the existence of the necessary \baup.

\begin{defn}
\label{def:balanced}
Let $\{w_i\}_{i=1}^m$ be a set of real weights, and let $M =
\max_{i\le m} |w_i|$ and $W = \sum_i w_i$. Let $\ell$ be a
non-negative integer. A permutation $\sigma \in S_m$ is
$\ell$-balanced, if for all $k$, $1 \le k \le m$,
\begin{equation}
\label{eq:l-balanced}
\min \{W,0\} - \ell M \le \sum_{i=1}^k w_{\sigma(i)} \le \max\{W,0\} +
\ell M \ .
\end{equation}
\end{defn}

\begin{defn}
\label{def:almost-uniform}
Let $\sigma$ be a random variable taking values in $S_m$, and let
$\alpha \in \real$. We call $\sigma$ a $\alpha$-uniform permutation if 
\[ \pr_{\sigma} [\sigma\{1,2,\dots,k\} = U ] \le \alpha \times {m \choose
k}^{-1}  \]
for every $k$, $1 \le k \le m$, and every $U \subseteq
\{1,2,\dots,m\}$ of size $k$.
\end{defn}

The main theorem from \cite{MS99} on the existence of \baup\ is the
following:

\begin{theorem}[\cite{MS99}]
\label{thm:baup}
There is a universal constant $C$ such that for any $m$ and any set
of weights $\{w_i\}_{i=1}^m$, there exists a $7$-balanced
$Cm^2$-uniform permutation on $\{w_i\}$. Moreover, if $|\sum_i w_i| > 15\max_i
|w_i|$, then there exists a $0$-balanced $Cm^2$-uniform permutation on
$\{w_i\}$.
\end{theorem}

\subsection*{Constructing a good flow}

\begin{lemma}
\label{lem:knapsack}
For arbitrary weights $\{a_i\}$ and $b$, there exists a multicommodity
flow $f$ in $G(\M_K)$ which routes one unit of flow between every pair
of unequal vertices, with $C(f) = O(|\Omega|n^5)$ and $L(f) = O(n)$.
\end{lemma}

\noindent Combining with Equation (\ref{eq:mixing-time-mk}) we
therefore conclude
\begin{theorem}[\cite{MS99}]
\label{thm:knapsack}
The mixing time of the Markov chain $\M_K$ satisfies $\tau_X(\eps) =
O(n^8 \ln \eps^{-1})$ for every starting state $X$. The chain is thus
rapidly mixing.
\end{theorem}

\smallskip \noindent {\bf Proof of Lemma~\ref{lem:knapsack}:} Let
$X,Y$ be arbitrary states of $\Omega$, $X \neq Y$. We wish to send one
unit of flow from $X$ to $Y$. As discussed earlier, our idea is to
spread this flow evenly among a family of \baup\ of $X \xor Y$, except
that we isolate a constant number of ``heavy'' items $H$ from $X \xor
Y$, and route the flow along \baup\ of $(X \xor Y)\setminus H$ and add
or remove
some elements of $H$ repeatedly along the path to maintain fine
balance (we always want the knapsack to be filled to capacity between
(roughly) $\min\{a(X),a(Y)\}$ and $\max\{a(X),a(Y)\}$: an upper bound
on the weight packed in the knapsack is clearly necessary to define a
feasible path, while the lower bound is used in the analysis to bound
the total flow through any edge by ``encoding'' each flow path which
uses that edge using an element of the state space).

We now proceed with the formal analysis. We wish to obtain an upper bound
on the maximum flow that passes through any state $Z$ (this will
clearly also provide an upper bound on the flow through any transition
$(Z,Z_1)$ of the chain). Let $X,Y$ be states such that the flow
between them passes through $Z$. Let $H$ be the $29$ elements of $X
\xor Y$ with the largest weight (set $H = X \xor Y$ if $|X \xor Y| \le
29$); breaking ties {\em according to index order}. Define $H_X = H
\cap X$, $H_Y = H \cap Y$, $S = (X \xor Y)\setminus H$ and $m=|S|$. Let
$\{w_i\}_{i=1}^m$ be an arbitrary enumeration of the weights of items
in $S$, where elements in $Y$ receive positive signs and those in $X$
receive in negative signs (since we want to add elements in $S \cap Y$
and remove those in $S \cap X$). The paths we use for our flow will
correspond to permutations of indices in $S$ that satisfy the specific
``balance'' requirement described below.

\medskip \noindent {\sf Claim.} {\em There is an absolute constant $C$
such that there exists a $Cm^2$-uniform family of permutations each
one (call it $\sigma$) of which satisfies the following ``balance''
condition:
\begin{equation}
\label{eq:balance-cond}
\min\{a(Y)-a(X),0\} - a(H_Y) \le \sum_{i=1}^k w_{\sigma(i)} \le
\max\{a(Y)-a(X),0\} + a(H_X), 
\end{equation}
for every $k$, $1 \le k \le m$.}

\medskip \noindent {\sf Proof.} We will assume $|X \xor Y| \ge 29$,
for otherwise $S = \emptyset$ and $m=0$, and there is nothing to prove.  Let $W =
\sum_{i=1}^m w_i = a(Y)-a(X)+a(H_X)-a(H_Y)$, and $M = \max_i
|w_i|$. Let us assume, w.l.o.g, that $W \ge a(H_X) - a(H_Y)$ (the
other case is symmetric), so it is easy to see that the above
condition (\ref{eq:balance-cond}) is equivalent to
\begin{equation}
\label{eq:ti-1}
- a(H_Y) \le \sum_{i=1}^k w_{\sigma(i)} \le W + a(H_Y)
\end{equation}
Comparing with condition (\ref{eq:l-balanced}), and allowing for both
cases $W \ge 0$ and $W < 0$, it is easy to see that an $\ell$-balanced
permutation satisfies (\ref{eq:ti-1}) above whenever $\ell M \le
\min\{a(H_Y),W+a(H_Y)\}$. Thus, when $|W| > 15M$, we can use
$0$-balanced permutations guaranteed by Theorem~\ref{thm:baup} for our
purposes. When $|W| \le 15M$, we have $a(H_X) - a(H_Y) \le W \le 15M$.
Also $a(H_X) + a(H_Y) = a(H) \ge 29 M$. Combining these two
inequalities we get $a(H_Y) \ge 7M$. Thus when $W \ge 0$, we have $7M
\le \min\{a(H_Y),W+a(H_Y)\}$, and thus we can use a $Cm^2$-uniform
family of $7$-balanced permutations to satisfy (\ref{eq:ti-1}). When
$W < 0$, we have $W \ge -15M$ and together with $a(H_X) - a(H_Y) \le
W$ this implies
\[ a(H_Y) \ge \frac{29 M - W}{2} \ge \frac{14 M - 2 W}{2} = 7 M - W \]
and thus once again $7M \le \min\{a(H_Y),W+a(H_Y)\}$, and we can use a
$7$-balanced $Cm^2$-uniform family of permutations. {\hfill $\Box$
{\em (Claim)}}

\medskip \noindent We now specify the flow paths between $X$ and $Y$
(the flow will be evenly split among all these paths). The paths will
follow the permutations $\sigma$ of the family guaranteed by above
Claim, except that along the way we will use elements of $H$ to keep
the knapsack as full as possible, and we will remove elements of $H$
as necessary to make room for elements of $S \cap Y$ to be
added. Hence each intermediate state will be of the form $H_0 \cup
((X\setminus H_X) \xor \{\sigma(1),\dots,\sigma(k)\})$ for some $k \le
m$ and $H_0 \subseteq H$. The path corresponding to a particular
$\sigma$ is defined by the following transitions:
\begin{itemize}
\item If $k < m$ and $w_{\sigma(k+1)} > 0$, then add $\sigma(k+1)$ if
possible (i.e., current knapsack has room for the item); else delete
an (arbitrary) element from $H_0$.
\item If $k < m$ and $w_{\sigma(k+1)} < 0$, then add an element form
$H - H_0$ if possible (so that knapsack is near full); otherwise remove
$\sigma(k+1)$. 
\item If $k=m$ (i.e., all elements in $S$ have been handled), add an
element of $H_Y$ if possible; otherwise delete an element from $H_X$.
\end{itemize}
By the upper bound of Condition (\ref{eq:balance-cond}), we have $a(X)
- a(H_X) + \sum_{i=1}^{k+1} w_{\sigma(i)} \le \max\{a(X),a(Y)\} \le b$ so
that we can always remove enough elements of $H$ to make room for
$w_{\sigma(k+1)}$ during its turn to be added. Moreover, the lower
bound of Condition (\ref{eq:balance-cond}) implies that for any
intermediate state $Z$ on any flow path, $a(Z \cup H) \ge
\min\{a(X),a(Y)\}$, and since we always keep the knapsack as full as
possible, there exist elements $h_1,h_2 \in H$ such that $a(Z \cup
\{h_1,h_2\}) \ge \min\{a(X),a(Y)\}$. In what follows $h_1,h_2$ are
fixed elements of $H$ that depend only on $Z,X,Y$.

To estimate the flow through $Z$, we will ``encode'' each pair $X,Y$
of states whose flow paths use $Z$ by a state $Z' \in \Omega$ (plus
some auxiliary information), so that we can argue that $C(f)$ is not
too large compared to $|\Omega|$. The encoding $Z'$ is defined by
\[ Z' = \big((X \xor Y) \setminus (Z \cup \{h_1,h_2\})\big) \cup (X \cap Y). \]
(Note that this is the complement of $Z \cup \{h_1,h_2\}$ in the
multiset $X \cup Y$. Thus it is reasonable to expect that $Z'$ will
supply a lot of the ``missing'' information about $X,Y$ that cannot be
obtained from $Z,h_1,h_2$.) Now
\begin{eqnarray*}
a(Z') & = & a(X) + a(Y) - a(Z \cup \{h_1,h_2\}) \\
& \le & a(X) + a(Y) - \min \{a(X),a(Y) \} \\
& = & \max\{a(X),a(Y)\} \le b
\end{eqnarray*}
so that $Z' \in \Omega$.

We now wish to upper bound the number of pairs $(X,Y)$ that could be
mapped to a given $Z'$.  Note that $Z \cap Z' = X \cap Y$ and $Z' \xor
(Z \cup \{h_1,h_2\}) = X \xor Y$, and knowing $X \xor Y$, we also know
$H$ (since these form the $29$ largest elements of $X \xor Y$, ties
broken according to index order). Thus $Z,Z',h_1,h_2$ together fix $X
\cap Y$, $X \xor Y$, $H$ and $S = (X \xor Y)\setminus H$. In order to
completely specify $X$ and $Y$, we add some more information to the
encoding, namely the subset $U \subseteq S$ that have been
``affected'' (i.e., added/removed) by the time the path from $X$ to
$Y$ reaches $Z$, and also $H' = H \cap X$.\footnote{The encoding $U$
we use is slightly different from the one Morris and
Sinclair~\cite{MS99} use in their proof.}  Thus, the pair $(X,Y)$ one
of whose flow paths passes through $Z$ is encoded by the $5$-tuple:
\[ f_Z(X,Y) = (Z',h_1, h_2, U, H') \ . \]
We now verify that $Z$ and $f_Z(X,Y)$ do pinpoint $X,Y$. Indeed, we
already argued $Z,Z',h_1,h_2$ alone fix $X \cap Y$, $X \xor Y$, $H$
and $S$. Now it is easy to verify that $X = (U \cap Z') \cup ((S
\setminus U) \cap Z) \cup (X \cap Y) \cup H'$ and similarly $Y = (U
\cap Z) \cup ((S \setminus U) \cap Z') \cup (X \cap Y) \cup
(H\setminus H')$.

We are now ready to bound $C(f)$ by estimating the cumulative flow
$f(Z)$ through $Z$. For each $X,Y$ such that there is a flow path from
$X$ to $Y$ passing through $Z$ and whose encoding equals $f_Z(X,Y) =
(Z',h_1, h_2, U, H')$, there will be non-zero flow only for paths
corresponding to those permutations $\sigma$ of $\{1,2,\dots,m\}$
(here $m =|S|$) that satisfy $\sigma\{1,2,\dots,|U|\} = U$. By the
$Cm^2$-uniformity of the family of permutations we use to spread the
flow, we can conclude that the total flow over all such paths is at
most $Cm^2 {m \choose |U|}^{-1}$. Thus summing over all $U \subseteq
S$, we still have only $Cm^3$ units of flow for each fixed
$(Z',h_1,h_2,H')$. Now there are $|\Omega|$ choices for $Z'$, and
$n^2$ choices for the pair $(h_1,h_2)$, and once $(Z',h_1,h_2)$ are
fixed, so is $H$, and thus there are at most $2^{29}$ possible choices
of $H' \subseteq H$ for each choice of $Z'$. In all, we have
\[ f(Z) \le |\Omega| \cdot n^2 \cdot 2^{29} \cdot Cm^3 = O(|\Omega|
n^5) \ . \]
Thus $C(f) = O(|\Omega| n^5)$ as well, and since all paths we
use to route flows clearly have length $O(n)$, $L(f) = O(n)$, and the
proof of Lemma~\ref{lem:knapsack} is complete. {\hfill $\Box$}

\section{Coupling and Path Coupling}
\label{sec:coupling}

We have so far focused on conductance based techniques for proving
rapid mixing, and saw a non-trivial application to sampling $0$-$1$
knapsack solutions. The classical approach to bounding the mixing time
is in fact via a different approach, viz. {\em Coupling}. The basic
idea behind the coupling argument is very intuitive: suppose we wish
to show that a Markov chain $\M$ starting from distribution $\pi'$
converges to its stationary distribution $\pi$ within a small number
of steps. Consider running the chain on a {\em joint} process $({\cal
X},{\cal Y})$ where both ${\cal X},{\cal Y}$ are individually {\em
faithful} copies of $\M$ and where ${\cal X}$ starts of at state $X_0$
distributed according to $\pi'$ and ${\cal Y}$ starts of in state
$Y_0$ distributed according to $\pi$.  Thus at any time step $t$, the
distribution of $Y_t$ equals $\pi$. Now if the joint evolution of
$(X_t,Y_t)$ is designed to encourage them to {\em coalesce} rapidly,
i.e., the ``distance'' between $X_t$ and $Y_t$ decreases rapidly, then
for large enough $t$, say $t \ge t'$, we will have $X_t = Y_t$, with
high probability, say $(1-\eps)$. Since the distribution of $Y_t$ is
$\pi$, it is easy to see that this implies that the
mixing time to get within $\eps$ of the stationary distribution when
the chain starts off in distribution $\pi'$, is at
most $t'$ (by the ``Coupling Lemma'' which we will state and prove
formally shortly). 

\subsection{The Coupling Lemma}

\begin{defn}[Coupling]
\label{defn:coupling}
Let $\M$ be a finite, ergodic Markov chain defined on state space
$\Omega$ with transition probabilities $P(\cdot,\cdot)$. A (causal)
{\em coupling} is a joint process $({\cal X},{\cal Y})=(X_t,Y_t)$ on $\Omega \times
\Omega$, such that each of the processes ${\cal X}$, ${\cal Y}$, considered
marginally, is a faithful copy of $\M$. In other words, we require
that, for all $x,x',y,y' \in \Omega$,
\[ \pr [ X_{t+1} = x' | X_t = x \wedge Y_t = y ] = P(x,x') \quad
\mbox{and,} \]
\[ \pr [ Y_{t+1} = y' | X_t = x \wedge Y_t = y ] = P(y,y') \
. \qquad \Box \]
\end{defn}

Note that the above conditions are consistent with $(X_t)$ and $(Y_t)$
being independent evolutions of $\M$, but does not imply it. In fact the
whole point of Coupling is to allow for the possibility that
\[  \pr [ X_{t+1} = x' \wedge Y_{t+1} = y' | X_t = x \wedge Y_t = y ]
\neq  P(x,x') P(y,y') \]
in order to encourage $X_t$ and $Y_t$ to {\em coalesce} rapidly.

\medskip \noindent {\bf Remark.} In applications to bounding mixing
time, $(X_t)$ will typically be Markovian, while we allow ${\cal Y}$
to be Non-Markovian or history dependent, i.e., $Y_t$ could depend
upon $X_0,\dots X_t$ and $Y_0,\dots,Y_{t-1}$, as long as it remains
faithful to the original chain $\M$. One can also imagine allowing the
process ${\cal Y}$ to make its moves dependent on future moves of
${\cal X}$, i.e., $Y_t$ can depend upon $X_{t+1},X_{t+2}$, etc. Such a
coupling is called a {\em non-causal} coupling. We will only be
concerned with causal couplings here, and the term ``Coupling'' will
always refer only to a causal coupling.

\medskip If it can be arranged that coalescence occurs rapidly,
independently of the initial states $X_0,Y_0$, we may then deduce that
$\M$ is rapidly mixing. The key result here is the {\em Coupling
Lemma}, which seems to have first explicitly appeared in
\cite{aldous}.

\begin{lemma}[Coupling Lemma]
\label{coupling-lemma}
Let $\M$ be a finite, ergodic Markov chain, and let $(X_t,Y_t)$ be a
coupling for $\M$. Suppose that $\pr [X_t \neq Y_t ] \le \eps$,
uniformly over the choice of initial state $(X_0,Y_0)$. Then the
mixing time $\tau(\eps)$ of $\M$ (starting from any state) is bounded
above by $t$.
\end{lemma}
\proof Let $X_0 = x$ be arbitrary and let $Y_0$ be distributed
according to the stationary distribution $\pi$ of $\M$. Let $A
\subseteq \Omega$ be an arbitrary event. We have
\begin{eqnarray*}
\pr [ X_t \in A ] & \ge & \pr [ Y_t \in A \wedge X_t = Y_t ] \\
& \ge & 1 - \pr [Y_t \notin A] - \pr [X_t \neq Y_t] \\
& \ge & \pr[Y_t \in A ] - \eps \\
& = & \pi(A) - \eps,
\end{eqnarray*}
and this implies the variation distance between $P^t(x,\cdot)$ and
$\pi$, $\Delta_x(t)$, is at most $\eps$, as desired. {\hfill $\Box$}

\medskip \noindent In light of the above Lemma, Coupling is a natural
technique to prove rapid mixing of Markov chains. And as we will
convince the reader in this section and the next, Coupling is a very
crisp and elegant technique and when it works, it invariably
establishes better bounds on mixing time than known through
conductance, and avoids the slackness which is typical of
conductance/canonical paths based proofs. We illustrate this by a
simple example below.

\subsection{An illustrative example of Coupling in action}
\label{sec:k-sets}

We consider the ``Bernoulli-Laplace diffusion model'', whose state
space $\Omega$ is the set of all $k$-element subsets of $[n] =
\{1,2,\dots,n\}$, and we wish to sample an element u.a.r from
$\Omega$. We assume $k \le n/2$ without loss of generality. A natural
chain on $\Omega$ is the following (let the current state be the
subset $X \subseteq [n]$ with $|X|=k$)
\begin{itemize}
\item Pick $r_X \in \{0,1\}$ u.a.r; If $r_X =0$, remain at $X$.
\item If $r_X = 1$, pick $i \in X$ u.a.r and $j \in [n] \setminus X$
u.a.r and move to $Y = X \cup \{j\} \setminus \{i\}$.
\end{itemize}

\noindent It is easy to that this chain is ergodic with uniform
stationary distribution $\pi(X) = N^{-1}$ for all $X \in \Omega$,
where $N = {n \choose k}$. We will show using Coupling that this chain
mixes in $O(k \log (k/\eps))$ time (we will later mention the sort of
weak bounds that more complicated conductance/resistance based proofs
give even for this very simple example).

\begin{theorem}
\label{thm:coupling-appl-1}
The mixing time of the above Markov chain satisfies $\tau_X(\eps) =
O(k \log (k/\eps))$ irrespective of the starting state $X$.
\end{theorem}
\proof The proof is based on a Coupling that is actually quite simple
to set up. The transition $(X_t,Y_t) \rightarrow (X_{t+1},Y_{t+1})$ is
defined as follows:
\begin{enumerate}
\item If $X_t = Y_t$, then pick $X_{t+1}$ as $\M$ would and set
$Y_{t+1} = X_{t+1}$; else
\item If $r_{X_t} = 0$, set $X_{t+1} = X_t$, and $Y_{t+1}=Y_t$.
\item \label{enum:op} If $r_{X_t} = 1$, then: Let $S = X_t \setminus Y_t$ and
$T = Y_t \setminus X_t$ (note that $|S| = |T|$); fix an arbitrary bijection $g
: S \rightarrow T$. Pick $i \in X_t$ u.a.r and $j \in [n]\setminus
X_t$ u.a.r and set $X_{t+1} = X_t \cup \{j\} \setminus \{i\}$. Define
$i' \in Y_t$ and $j' \in [n]\setminus Y_t$ as follows:
\begin{itemize}
\item If $i \in X_t \cap Y_t$, then $i'=i$, else $i' = g(i)$
\item If $j \notin Y_t$, $j'=j$, else (now $j \in T$) $j' = g^{-1}(j)$.
\end{itemize}
Now set $Y_{t+1} = Y_t \cup \{j'\} \setminus \{i'\}$. 
\end{enumerate}

It is easy to see that $(X_t)$ and $(Y_t)$ are individually just
copies of $\M$, so the above is a legal (in fact Markovian) coupling.
We assume $k \ge 2$ to avoid trivialities. Denote by $D_t$ the
random variable $X_t \xor Y_t$. We wish to bound the expectation
\begin{equation}
\label{eq:jki}
\av [ |D_{t+1}| | D_t ] \le (1 - \frac{1}{k}) |D_t| \ ,
\end{equation}
as this will imply $\av [ |D_t| | D_0 ] \le (1-\frac{1}{k})^t
|D_0|$. 
Since $|D_t|$ is a non-negative integer random variable, and $|D_0|
\le 2k$, we obtain
\begin{eqnarray*}
\pr [ |D_t| > 0 | D_0 ] & \le & \av [ |D_t| | D_0 ] \\
& \le & 2k \cdot (1-\frac{1}{k})^t 
\end{eqnarray*}
which is at most $\eps$ provided $t \ge k \ln (2k\eps^{-1})$.
Invoking the Coupling Lemma~\ref{coupling-lemma}, we obtain that the
mixing time is $O(k \ln (k/\eps))$, as promised. It remains therefore
to establish (\ref{eq:jki}) which basically quantifies the fact that
$X_t$ and $Y_t$ tend to ``coalesce''.

Let $q = |X_t \xor Y_t|$, and let $q' = |X_{t+1} \xor Y_{t+1}|$. We
want the expectation of $q'$ for a given $q$. Consider now the choices
in Step (\ref{enum:op}) of the Coupling. Four cases now arise:
\begin{itemize}
\item[(i)] $j \in [n] \setminus(X_t \cup Y_t)$ and $i \in X_t
\setminus Y_t$: Then $q'=q-2$.
\item[(ii)] $j \in Y_t \setminus X_t$ and $i \in X_t \cap Y_t$: Then
$q' = q-2$.
\item[(iii)] $j \in  Y_t \setminus X_t$ and $i \in X_t \setminus Y_t$,
$j \neq g(i)$: Then $q' = q-4$.
\item[(iv)] In all other cases $q'=q$.
\end{itemize}

Thus the expected value of the change $q'-q$ is
\[ \av [ q'-q ] = \frac{n-k-q/2}{n-k} \cdot \frac{q/2}{k} \cdot (-2)+
\frac{q/2}{n-k}\cdot\frac{k-q/2}{k} \cdot (-2) +
\frac{q/2}{n-k}\cdot\frac{q/2-1}{k} \cdot (-4) \]
and this gives $\av [q'|q] \le (1 - \frac{n-2}{k(n-k)}) q \le
(1-\frac{1}{k})q$ (as $k \ge 2$).  {\hfill $\Box$ {\em
(Theorem~\ref{thm:coupling-appl-1})}}

\bigskip \noindent {\bf Comparison with performance of Canonical
Paths.} The best bound achievable for this problem via the canonical
paths/conductance based approach seems to be (see \cite{sinclair92})
to bound $\bar{\res}$ by demonstrating a fractional flow that routes
one unit between every pair of unequal states, and this gives
$\bar{\res} \le \frac{k^2(n-k)^2}{n(n-1)}$. For $k = \Omega(n)$, say
$k=n/2$, this gives a bound on mixing time equal to $O(n^2\log ({n
\choose k} \eps^{-1})) = O(n^3 + n^2 \log \eps^{-1})$, which is
significantly worse than the $O(n (\log n + \log \eps^{-1}))$ bound we
proved using Coupling! In fact, in this case ($k=n/2$), the
second-largest eigenvalue is known exactly: $\lambda_1 = 1 - 2/n$, so
that even getting the best bound on the spectral gap, only yields a
mixing time of $O(n^2 + n \log \eps^{-1})$ (using
Proposition~\ref{prop:spectrum})! These crisp and significantly
improved bounds seem to be typical of Coupling whenever it works.  We
will later (in Section~\ref{sec:uni-knap}) also see an application of
sampling from subsets of $[n]$ of size {\em at most} $k$ (this is just the
``uniform'' version of the knapsack problem, where all items have the
same size), where Coupling gives a much better bound mixing time than
seems possible using techniques of Section~\ref{sec:knapsack-MS}.

\subsection{Known applications of Coupling}
\label{subsec:coupling-appl}

Owing to its intuitive appeal, Coupling has been a very popular and
successful technique in rapid mixing results. Some instructive
examples of Coupling that have appeared in the literature are in
sampling proper $k$-colorings of a
graph~\cite{jerrum-coloring,BD97,vigoda}, linear extensions of a
partial order~\cite{mat91,BD-linear-extensions}, points in a convex
body~\cite{BDJ-convex}, independent sets in low-degree
graphs~\cite{LV97,BD97,DG-indset}, general contingency tables on $2$
rows~\cite{DG-contingency}, etc. Even Broder's original
paper~\cite{broder} on sampling from the set of perfect matchings of a
bipartite graph used a complicated Coupling argument, which was later
found to have an error~\cite{mihail-broder}.

\subsection{Path Coupling}
\label{subsec:path-coupling}

Despite the conceptual simplicity and appeal of Coupling, it can often
get very difficult to design couplings appropriate to specific
situations that arise in sampling problems. The problem is one of
``engineering'': how do we encourage $(X_t)$ and $(Y_t)$ to coalesce
while at the same time meeting the apparently contradicting
requirement of keeping the individual processes faithful to $\M$?
This can lead to severe technical complexities (see \cite{mat91} to
get an impression of this). This led Bubley and Dyer~\cite{BD97} to
invent an elegant solution to the task of designing Couplings: they
called it ``Path Coupling''. The idea behind Path Coupling is to
define the coupling only for ``adjacent'' states, i.e., only for pairs
of states in a carefully chosen subset $S$ of $\Omega \times \Omega$
(and hopefully the task is easier for such pairs of states), and then
extend the coupling to arbitrary pairs of states by composition of
adjacent couplings along a path. In fact, the discovery of Path
Coupling has led to a spurt of Coupling based rapid mixing proofs, and
indeed most of the applications cited in
Section~\ref{subsec:coupling-appl} use Path Coupling. We now state and
prove the ``Path Coupling'' lemma (a version taken from
\cite{DG-coloring}):

\begin{lemma}[Path Coupling Lemma]
\label{path-coupling-lemma}
Let $\delta$ be an integer valued metric defined on $\Omega \times
\Omega$ which takes values in $\{0,1,\dots,D\}$. Let $S$ be a subset
of $\Omega \times \Omega$ such that for all $(X_t,Y_t) \in \Omega
\times \Omega$, there exists a path $X_t = Z_0, Z_1,\dots, Z_r = Y_t$
between $X_t$ and $Y_t$ where $(Z_{\ell},Z_{\ell+1}) \in S$ for $0 \le
\ell < r$, and $\sum_{\ell=0}^{r-1} \delta(Z_{\ell},Z_{\ell+1}) =
\delta(X_t,Y_t)$. (Equivalently, $\delta$ is defined by specifying a
graph $H$ with vertex set $\Omega$ and edge set $S$, and weights on
edges in $S$, and $\delta(X,Y)$ is simply the shortest path between
$X$ and $Y$ in this graph.) Suppose a Coupling $(X,Y) \mapsto (X',Y')$
of the Markov chain $\M$ is defined on all pairs $(X,Y) \in S$ (note
that $(X',Y')$ need not lie in $S$) such that there exists a $\beta <
1$ such that $\av [ \delta(X',Y')] \le \beta \av [ \delta(X,Y) ] $ for
all $(X,Y) \in S$. Then the mixing time $\tau(\eps)$ of $\M$ satisfies
$\tau(\eps) \le \frac{\ln (D \eps^{-1})}{(1-\beta)}$.
\end{lemma}

\smallskip \noindent {\bf Remark.} One can also bound the mixing time
in the case $\beta = 1$~\cite{BD97,DG-coloring}. For the
applications we will use to illustrate this technique, we will
actually have $\beta < 1$, so to keep things simple we do not discuss
the $\beta = 1$ case.

\medskip \proof First, we observe that the Coupling on $S$ can be
extended in an obvious way to a Coupling on the entire space $\Omega
\times \Omega$. Indeed let $(X_t,Y_t) \in \Omega \times \Omega$. Pick
a ``path'' $X_t= Z_0, Z_1, \dots, Z_r = Y_t$ such that $\delta(X,Y) =
\sum_{\ell =0}^{r-1} \delta(Z_{\ell},Z_{\ell+1})$ (use a deterministic
choice rule for resolving ties). Define the coupling $(X_t,Y_t)
\mapsto (X_{t+1},Y_{t+1})$ as follows: First select $X_{t+1} = Z'_0
\in \Omega$ according to the probability distribution
$P(X,\cdot)$. Now select $Z'_1$ according to the distribution induced
by the pairwise coupling of the adjacent states $Z_0$ and $Z_1$,
conditioned on the choice of $Z'_0$; then select $Z'_2$ using the
pairwise coupling of $(Z_1,Z_2)$, and so on, ending with $Z'_{r} =
Y_{t+1}$. It is easy to verify, by induction of the path length $r$,
that $Y_{t+1}$ has been selected according to the distribution
$P(Y_t,\cdot)$, so $(X_t,Y_t) \mapsto (X_{t+1},Y_{t+1})$ does define a
``legal'' coupling that obeys conditions of
Definition~\ref{defn:coupling}. Now
\begin{eqnarray*}
\av [ \delta(X_{t+1},Y_{t+1}) ] & \le & \av [ \sum_{\ell =0}^{r-1}
\delta(Z'_{\ell},Z'_{\ell+1}) ]\\
& = & \sum_{\ell=0}^{r-1} \av [ \delta(Z'_{\ell},Z'_{\ell+1}) ] \\
& \le & \beta \sum_{\ell=0}^{r-1} \delta(Z_{\ell},Z_{\ell+1}) \\
& = & \beta \delta(X_t,Y_t),
\end{eqnarray*}
where we have used the fact that $\delta$ is a metric, and linearity
of expectation. Now as in the proof of
Theorem~\ref{thm:coupling-appl-1}, this gives $\av [ \delta(X_t,Y_t) ]
\le \beta^t D$, and thus $\pr [ X_t \neq Y_t] \le \av [ \delta(X_t,Y_t)
] \le \eps$ whenever $t \ge \frac{\ln(D\eps^{-1})}{(1-\beta)}$. Invoking
the Coupling Lemma~\ref{coupling-lemma}, the claimed bound on the
mixing time $\tau(\eps)$ follows. {\hfill $\Box$}

\medskip \noindent {\bf Remark.} The notion of ``adjacency'' in the
graph $H$ defined for Path Coupling need not have anything to do with
the transitions in the Markov chain that is being studied. In fact, two
states that are adjacent in the Path Coupling graph $H$ need not even
be reachable from one another in the Markov chain.

\section{Some applications of Path coupling}
\label{sec:path-coupling-applns}

In this section, we present a few applications of path coupling to
Markov chains for interesting sampling problems. 

\subsection{Sampling $k$-colorings of a graph}
\label{subsec:pc-coloring}
Given a graph $G=(V,E)$ with maximum degree $\Delta$, consider the task of
sampling uniformly at random from the set $\Omega_k(G)$ of (proper)
$k$-colorings of $G$. Let $C=\{1,2,\dots,k\}$ be the set of colors.

A natural Markov chain for the above problem, known in the literature
as ``Glauber dynamics'', is the following. Suppose the current state
is a coloring $X$:
\begin{itemize}
\item Choose $v \in V$ u.a.r, and $c \in C$ u.a.r. If $X_{v
\rightarrow c}$ (i.e., $X$ with the color of $v$ changed to $c$) is a
proper coloring of $F$, then move to $X_{v\rightarrow c}$, else remain
at $X$.
\end{itemize}

Jerrum~\cite{jerrum-coloring} (see also \cite{jerrum-survey}) first
proved, using Coupling, that the above chain rapidly mixes for $k > 2
\Delta$. We will now present a simple proof due to Bubley and
Dyer~\cite{BD97} of this fact using Path Coupling. We remark that
Vigoda~\cite{vigoda} recently established that this chain mixes
rapidly for $k > \frac{11}{6} \Delta$, using Path Coupling on a different
chain, and then using that to deduce the mixing time of the Glauber
dynamics. 

Following \cite{BD97}, we present the result in a more general set-up
that captures ``coloring-type'' problems, and then deduce the result
for coloring from that. 

The general set-up is the following. Let $V$ and $C$ be finite sets,
and let $n = |V|$ and $k = |C|$, and we consider a finite Markov chain
$\M$ with state space $\Omega \subseteq C^V$, the set of functions
from $V$ to $C$, and unique stationary distribution $\pi$. The
transition structure of $\M$ is similar to the graph coloring case we
considered above: From a current state $X \in \Omega$, pick $v \in V$
according to a fixed distribution $J$ on $V$, and
and $c \in C$ according to a distribution $\kap_{X,v}$ that depends only
on $X$ and $v$, and make the transition to $X_{v\rightarrow c}$
(where $X_{v\rightarrow c}(w)$ equals $c$ if $w = v$, and equals
$X(w)$ otherwise). We assume that $\kap_{X,v}(c) = 0$ whenever $X_{v
\rightarrow c} \notin \Omega$. Path Coupling yields the following
result for this class of problems (for distributions $A$ and $B$, $\|
A - B \|$ denotes their statistical difference or variation distance):

\begin{theorem}[\cite{BD97}]
\label{thm:bubley-path-coupling}
Let $\Omega = C^V$, and let
\[ \beta = \max_{X,Y \in \Omega, i \in V} \Big\{ 1 - J(i) + \sum_{j
\in V} J(j) \| \kap_{X,j}-\kap_{Y,j} \| ~~ | ~~ Y = X_{i \ra c} \mbox{ for
some } c \in C, \mbox{ and } Y \neq X \Big\}\ . \]
Then, if $\beta < 1$, the mixing time of $\M$ satisfies
$\tau(\eps) \le \ln (n\eps^{-1})/(1-\beta)$. 
\end{theorem}
\proof We set up a Path Coupling with ``adjacency graph'' being all
non-equal pairs $(X,Y)$ such that $Y = X_{i \ra c}$ for some $i,c$,
and the metric $\delta$ used is the Hamming metric (so $\delta(X,Y) =
1$ for adjacent pairs). For such a pair $(X,Y)$ define the coupling to
$(X',Y')$ as follows: $X'$ is distributed according to $P(X,\cdot)$,
namely: pick $v \in V$ according to $J$ and $c_0 \in C$ according to
$\kap_{X,v}$, and set $X' = X_{v\ra c_0}$. Next pick $c_1 \in C$ as
follows: with probability $\min\{1,\kap_{Y,v}(c_0)/\kap_{X,v}(c_0)\}$
let $c_1 = c_0$, otherwise pick $c_1$ according to the distribution
$\gamma(c) = \frac{\max\{0,\kap_{Y,v}(c)-\kap_{X,v}(c)\}}{\|\kap_{Y,v} -
\kap_{X,v}\|}$.

It is easy to see that marginally we choose $c_1$ according to
$\kap_{Y,v}$, so the above defines a ``legal'' coupling for the chain
$\M$. It is also easy to verify that $\pr [ c_1 \neq c_0 ] = \|
\kap_{Y,v} - \kap_{X,v} \|$. Now since $\delta(X,Y)$ changes by at
most $1$ in one step of the chain, we have
\begin{eqnarray*}
\av [ \delta(X',Y') ] & = & 1 - \pr [ \delta(X',Y') = 0 ] + \pr
[\delta(X',Y')=2] \\
& = & 1 - J(i) \pr [ c_0 = c_1 | v = i ] + \sum_{j \neq i} J(j) \pr
[c_0 \neq c_1 | v = j ] \\
& = & 1 - J(i) (1-\|\kap_{Y,i} -\kap_{X,i} \|) + \sum_{j \neq i} J(j)
\| \kap_{Y,j} - \kap_{X,j} \| \\
& \le & \beta \delta(X,Y) 
\end{eqnarray*}
(since $\delta(X,Y) = 1$). The result now follows from the Path
Coupling Lemma~\ref{path-coupling-lemma}. {\hfill $\Box$}

\medskip \noindent {\bf Application to Coloring.} Consider the Markov
chain with state space all (not necessarily proper) $k$-colorings of
$G$ and transitions at state $X$ defined as follows.
\begin{enumerate}
\item Choose $v$ at random from $V$ according to distribution $J$ and
$c$ u.a.r from $C$.
\item If $v$ is properly colored in $X_{v\ra c}$, then move to $X' =
X_{v \ra c}$ else remain at $X$.
\end{enumerate}
This is an extension of the Glauber dynamics we discussed earlier
(except that we allow more general distributions to select $v$ from),
to all of $C^V$ (we do so in order to be able to apply
Theorem~\ref{thm:bubley-path-coupling}). This does not cause any
problems since the non-proper colorings are transient states, and the
stationary distribution is uniform over all proper $k$-colorings of
$G$, and zero elsewhere. Moreover, if we start from a proper
$k$-coloring, then we visit only states that correspond to proper
$k$-colorings, so the mixing time of this chain is an upper bound on
the mixing time of the Glauber dynamics. Note that this chain is not
reversible, but Theorem~\ref{thm:bubley-path-coupling} applies for
such chains as well.

Let us now apply Theorem~\ref{thm:bubley-path-coupling}. Let $d(v)$
denote the degree of vertex $v$, and let $m$ be the number of edges in
$G$. We will use $J$ to be proportional to the degree of the vertex,
so that $J(v) = d(v)/2m$. If colorings $X$ and $Y$ differ only on
vertex $i$, then $\kap_{Y,j} = \kap_{X,j}$ unless $j = i$ or $j \sim
i$ (here $j \sim i$ stands for adjacency in the graph $G$). When
$j=i$, $\kap_{X,i}(X(i)) = \frac{d(i)+1}{k}$ and $\kap_{Y,i}(X(i)) =
\frac{1}{k}$, and similarly for the color $Y(i)$, while $\kap_{X,i}(c)
= \kap_{Y,i}(c)$ for all colors $c \neq X(i),Y(i)$. Hence $\|
\kap_{Y,i} - \kap_{X,i} \| = d(i)/k$. When $j \sim i$, every color
that would be accepted in $X$ (resp. $Y$), except possibly $Y(i)$
(resp. $X(i)$) would be accepted in $Y$ (resp. $X$) as well, and hence $\|
\kap_{Y,j} - \kap_{X,j} \| = \frac{1}{k}$. Thus the parameter $\beta$
(from Theorem~\ref{thm:bubley-path-coupling}) satisfies
\[ \beta \le 1 - \frac{d(i)}{2m} (1 - \frac{d(i)}{k} ) +
\sum_{j \sim i} \frac{d(j)}{2mk} \ . \]
Hence $\beta < 1$ whenever
\[ k > \max_{v \in V} \{ d(v) + \sum_{w \sim
v} \frac{d(w)}{d(v)} \} \ . \]
This condition is certainly satisfied when $k > 2\Delta$, so using
Theorem~\ref{thm:bubley-path-coupling} we conclude
\begin{theorem}[\cite{BD97}]
\label{thm:glauber}
The Glauber dynamics for sampling proper $k$-colorings of a graph $G$
with maximum degree $\Delta$ is rapidly mixing (with mixing time $O(kn
\log (n\eps^{-1}))$) whenever $k > 2\Delta$.
\end{theorem}

\subsection{Sampling ``Uniform Knapsack'' solutions}
\label{sec:uni-knap}
We consider another elegant application of Path Coupling. We are
interested in sampling from the space $\Omega$ of subsets of $[n]
=\{1,2,\dots,n\}$ of size {\em at most} $k$. This resembles the
problem of sampling $k$-element subsets of $[n]$ that we considered in
Section~\ref{sec:k-sets}, but turns out to be trickier. Note also that
this problem is a special case of the $0$-$1$ knapsack problem (which
we considered in Section~\ref{sec:knapsack-MS}) when all items to be
packed have the same size. 

The Markov chain $\M_{K}$ we will study will be the same as the one in
Section~\ref{sec:knapsack-MS}, namely from a state $X \subseteq [n]$,
$|X| \le k$, pick $r_X \in \{0,1\}$ u.a.r. If $r_X = 0$ remain at $X$. If
$r_X=1$, pick an $i \in [n]$ u.a.r and move to $X \setminus \{i\}$ if
$i \in X$ and to $X \cup \{i\}$ if $i \notin X$ and $|X| < k$. We will
use Path Coupling to prove
\begin{theorem}
\label{thm:uni-knap}
The mixing time of the Markov chain $\M_{K}$ satisfies $\tau(\eps) =
O(n \log (k\eps^{-1}))$. 
\end{theorem}
\proof We will use Path Coupling with the (somewhat unusual) metric
$\delta(X,Y) = |X \xor Y| + | |X| - |Y| |$. Note that $\delta(X,Y) \ge
2$ whenever $X \neq Y$. The set of ``adjacent'' pairs $S \subseteq
\Omega \times \Omega$ for which we will define the Coupling is: $S =
\{(X,Y) : X,Y \in \Omega \wedge \delta(X,Y) = 2\}$. It is easy to see
that the metric $\delta$ and the set $S$ satisfy the conditions
required by the Path Coupling Lemma~\ref{path-coupling-lemma}. 

Now consider $(X,Y) \in S$ with $\delta(X,Y) = 2$; we wish to define
a Coupling $(X,Y) \mapsto (X',Y')$. There are two possibilities for
$(X,Y)$:
\begin{itemize}
\item[(i)] One of $X,Y$ is a subset of the other, say $Y \subset X$
(the other case is symmetric),
$|Y| = |X| - 1$.
\item[(ii)] $|X| = |Y|$ and $|X \xor Y| = 2$.
\end{itemize}
\noindent We consider each of these cases in turn. 

\medskip \noindent {\sf Case (i)}: Let $Y = X \setminus \{p\}$ for
some $p \in [n]$.  Now the Coupling $(X',Y')$ is defined as follows:
\begin{itemize}
\item[(1)] Pick $r_X \in \{0,1\}$ and $i \in [n]$ u.a.r. If $i =p$
then set $r_Y = 1-r_X$; otherwise set $r_Y = r_X$. 
\item[(2)] If $r_X = 0$ set $X'=X$. Else if $i \in X$ set $X' = X \setminus
\{i\}$, else set $X' = X \cup \{i\}$ if $|X| < k$ and $X'=X$ otherwise.
\item[(3)] If $r_Y = 0$ set $Y'=Y$. Else if $i \in Y$ set $Y' = Y \setminus
\{i\}$, else set $Y' = Y \cup \{i\}$ if $|Y| < k$ and $Y'=Y$ otherwise.
\end{itemize}
It is easy to see that $\delta(X',Y') = 2$ except when $i=p$, in which
case, since we have cleverly designed the Coupling by setting $r_Y =
1-r_X$ so that only one of $X,Y$ ``fires'', $\delta(X',Y') = 0$. Thus
we have $\av [ \delta(X',Y') ] = (1-\frac{1}{n}) \delta(X,Y)$.

\medskip \noindent {\sf Case (ii)}: $|X| = |Y|$ and $|X \xor
Y|=2$. Let $X = S \cup \{p\}$ and $Y = S \cup \{q\}$ for some $p \neq
q$. The Coupling $(X',Y')$ is defined as follows:
\begin{itemize}
\item[(1)] Pick $r_X \in \{0,1\}$ and $i \in [n]$ u.a.r. Set $r_Y =
r_X$.  If $i \notin \{p,q\}$, set $j=i$. If $i=p$ (resp. $q$) set
$j=q$ (resp. $p$).
\item[(2)] If $r_X = 0$ set $X'=X$. Else if $i \in X$ set $X' = X \setminus
\{i\}$, else set $X' = X \cup \{i\}$ if $|X| < k$ and $X'=X$ otherwise.
\item[(3)] If $r_Y = 0$ set $Y'=Y$. Else if $j \in Y$ set $Y' = Y \setminus
\{j\}$, else set $Y' = Y \cup \{j\}$ if $|Y| < k$ and $Y'=Y$ otherwise.
\end{itemize}
Once again, the Coupling has been constructed so that $\delta(X',Y') =
2$ whenever $i \notin \{p,q\}$; $\delta(X',Y')=0$ if $i=p$ and
$r_X=1$, and $\delta(X',Y') \le 2$ in all cases.  Thus we have $\av [
\delta(X',Y') ] \le (1-\frac{1}{2n}) \delta(X,Y)$.

Combining both the above cases we get $\av [ \delta(X',Y') ] \le
(1-\frac{1}{2n}) \delta(X,Y)$ always. Also, the maximum value $D$ of
$\delta(X_0,Y_0)$ over all pairs $(X_0,Y_0) \in \Omega \times \Omega$
is clearly $2k$. By Theorem~\ref{thm:bubley-path-coupling} therefore,
we have shown that $\M_K$ has mixing time $\tau(\eps) = O(n \log
(k\eps^{-1}))$, completing the proof. {\hfill $\Box$}

\medskip \noindent {\bf Comparison with Canonical paths.} Even for
this special case of $0$-$1$ knapsack, the best bound that we get
using the multicommodity flow based analysis of
Section~\ref{sec:knapsack-MS} (without any change) is only $O(n^{6})$,
and it is almost inconceivable that such an approach can hope to yield
a bound better than $O(n^3)$. Coupling gave us a much better $O(n \log
(k\eps^{-1}))$ bound, and the proof was in fact much easier than using
canonical paths!

\medskip \noindent {\bf Remark.} The uniformity of weights seems
critical to our argument above. The ``asymmetry'' created when items
have widely varying sizes seems to make it difficult for any natural
Coupling strategy to work.

\subsection{Linear extensions of a partial order}

We are given a partially ordered set $(P,\preceq)$ where $|P| = n$,
and we want to sample u.a.r from the space $\Omega$ of all linear
orders that extend $\preceq$. (A linear order extending $\preceq$ is a
permutation $a_1,a_2,\dots,a_n$ of the elements of $P$ such that $a_i
\preceq a_j$ implies $i \le j$.)

A natural Markov chain with uniform stationary distribution over
$\Omega$ was shown to be rapid mixing by Karzanov and Khachiyan via
conductance arguments that exploited the geometry of the
space~\cite{KK90}. Dyer and Frieze~\cite{DF-volume} improved the
conductance estimate, and hence the bound on the mixing time, of this
chain, and this gave a mixing time of $O(n^5 \log n + n^4 \log
\eps^{-1})$. 

In this section, we will sketch a chain $\M^J_{\rm le}$, which is a
slight variant of the chain discussed above, and show (using Path
Coupling) that it has a mixing time of $O(n^3 \log (n\eps^{-1}))$,
which significantly improves the best ``conductance based'' bound for
this problem.  The chain and its analysis are due to Bubley and
Dyer~\cite{BD-linear-extensions} (see also \cite{jerrum-survey} for an
exposition).

Actually this algorithm can be used to sample u.a.r from any set
$\Omega$ of permutations of elements of $P$ that satisfies the
following ``closure'' property: If $\sigma = (a_1,a_2,\dots,a_n) \in
\Omega$ and $\sigma \circ (i,j) =
(a_1,\dots,a_{i-1},a_j,a_{i+1},$ $\dots,a_{j-1},a_i,a_{j+1},\dots,a_n) \in
\Omega$ (i.e., the positions of $a_i$ and $a_j$ can be swapped and the
resulting permutation still lies in $\Omega$), then all permutations
which are obtained from $\sigma$ by placing $a_i$ and $a_j$ at
arbitrary positions in the interval $[i,j]$, also lie in $\Omega$.
Clearly the linear extensions of a partial order have this closure
property.

The transitions from one linear extension to another in the chain are
obtained by (pre)-composing with a random transposition $(p,p+1)$ (if
this yields a valid linear order); however, instead of selecting $p
\in [n-1]$ uniformly, $p$ is chosen according to a distribution $J$ on
$[n-1]$ that gives greater weight to values near the center of the
range. Formally, the chain $\M^J_{\rm le}$ is defined as follows. Let
the current state be $X_t$. Then the next state $X_{t+1}$ is defined
by the following random experiment:
\begin{itemize}
\item[(1)] Pick $p \in [n-1]$ according to the distribution $J$, and
$r \in \{0,1\}$ u.a.r
\item[(2)] If $r=1$ and $X_t \circ (p,p+1) \in \Omega$, then $X_{t+1}
= X_t \circ (p,p+1)$; otherwise $X_{t+1}=X_t$.
\end{itemize}

To use Path Coupling we need to specify an ``adjacency'' structure for
the state space $\Omega$. We say two states $g$ and $g'$ are adjacent
if $g' = g \circ (i,j)$ for some transposition $(i,j)$ with $1 \le i <
j \le n$, and the ``distance'' $\delta(g,g')$ in this case is defined
to be $j-i$. Since this distance is symmetric (i.e., $\delta(g,g') =
\delta(g',g)$), this adjacency structure yields a weighted, undirected
graph $H$ on vertex set $\Omega$. One can verify that the shortest
path between adjacent states $g,g'$ in $H$ is the direct one that uses
the edge $(g,g')$. We may thus extend $\delta$ to a metric on $\Omega$
by defining $\delta(g,h)$ for arbitrary states $g,h \in \Omega$ to be
the length of a shortest path from $g$ to $h$ in $H$, and all
conditions of the Path Coupling Lemma~\ref{path-coupling-lemma} are
now met. It remains to define a coupling $(g,h) \mapsto (g',h')$ for
adjacent states $g,h$ and then bound $\av [ \delta(g',h') ]$. 

The Coupling is defined as follows. Let $(g,h)$ be a pair of adjacent
states in $H$ and let $h = g \circ (i,j)$. Then the transition to
$(g',h')$ is defined by the following experiment:
\begin{itemize}
\item[(i)] Pick $p \in [n-1]$ according to distribution $J$, and $r_g
\in \{0,1\}$ u.a.r. If $j-i = 1$ and $p = i$, set $r_h = 1 -r_g$;
otherwise set $r_h = r_g$.
\item[(ii)] If $r_g = 1$ and $g \circ (p,p+1) \in \Omega$ then set $g'
= g \circ (p,p+1)$ else set $g' = g$.
\item[(iii)] If $r_h = 1$ and $h \circ (p,p+1) \in \Omega$ then set $h'
= h \circ (p,p+1)$ else set $h' = h$.
\end{itemize}

\begin{lemma}
\label{lem:lin-order-1}
For adjacent states $g$ and $h$, for a suitable choice of the
probability distribution $J$, we have 
\[ \av ~ [ \delta(g',h') ~ | ~ g,h ] \le \Big( 1 - \frac{6}{n^3-n} \Big)
\delta(g,h) \ . \] 
\end{lemma}

In light of Lemma~\ref{path-coupling-lemma}, this implies that the
mixing time of $\M^J_{\rm le}$ is $O(n^3 \log (n\eps^{-1}))$ (since
the ``diameter'' $D$ of the graph $H$ is easily seen to be at most
${n \choose 2}$). It thus only remains to prove
Lemma~\ref{lem:lin-order-1}. 

\medskip \noindent {\bf Proof of Lemma~\ref{lem:lin-order-1}:} We only
provide the skeleton of the proof; details can be found in
\cite{BD-linear-extensions}. When $h = g \circ (i,j)$, it is easy to
see that when $p \notin \{i-1,i,j-1,j\}$, we will have $h' = g' \circ
(i,j)$ and thus $\delta(g',h') = \delta(g,h) = j-i$. When $p = i-1$ or
$p=j$, it is again easily checked that $\av ~ [ \delta(g',h') ~ | ~ g
, h, p =i-1 \vee p =j ] \le \delta(g,h) + 1/2$.

The ``interesting case'' is when $p=i$ or $p=j-1$. These are
symmetric, so let us focus on the case $p=i$. There are two sub-cases:
$j-i = 1$ and $j-i \ge 2$. First, consider the case $j-i=1$. In this
case, we have made sure, by setting $r_h = 1 - r_g$, that only one of
$g$ or $h$ ``fires'' in the Coupling, and thus $g' = h'$ and therefore
$\delta(g',h') = 0$! In the case $j-i \ge 2$, by the ``closure''
property of $\Omega$ discussed earlier (this is the only place where
we use this closure property), we know both $g \circ (i,i+1)$, $h
\circ (i,i+1) \in \Omega$, thus either $r_X = r_Y =0$ and then
$\delta(g',h') = \delta(g,h)$, or $r_X = r_Y = 1$ and $\delta(g',h') =
j-i-1 =\delta(g,h) -1$. Hence $\delta(g',h')$ is less than
$\delta(g,h)$ in expectation.

Summing up, it follows from the above discussion that
\begin{equation}
\label{eq:rty}
\av ~ [ \delta(g',h') ~ | g,h ] \le \delta(g,h) -
\frac{-J(i-1)+J(i)+J(j-1)-J(j)}{2} \ . 
\end{equation}
Specializing the probability distribution $J(\cdot)$ to be $J(p)
\eqdef \zeta (p+1)(n-p-1)$ where $\zeta = 6/(n^3-n)$ is a normalizing
constant, and using $\delta(g,h) = j-i$, we get from (\ref{eq:rty})
that $\av ~ [ \delta(g',h') ] \le (1-\zeta) \delta(g,h)$. {\hfill
$\Box$ {\em (Lemma~\ref{lem:lin-order-1})}}

\section{Coupling is weaker than Conductance}
\label{sec:kumar-ramesh}

We have seen several Coupling based proofs in the last Section which
are not only extremely simple and elegant, but also end up giving
much better bounds on mixing time than known via conductance based
arguments. So, is Coupling the panacea as far as bounding mixing times
goes? In particular, is Coupling as powerful as conductance, and does
it {\em capture} rapid mixing exactly?

This fundamental question was unanswered for a long time until
recently when Kumar and Ramesh~\cite{KR99} proved the following
important result: For the famous Jerrum-Sinclair chain for sampling
perfect and near-perfect matchings, no Coupling argument can show
rapid mixing (the chain is known to be rapidly mixing using a
canonical paths argument~\cite{JS-permanent}). Hence Coupling is
actually ``weaker'' than conductance! We discuss the salient features
behind their proof in this section.

\medskip \noindent {\bf The Jerrum-Sinclair Chain.} We are given a
bipartite graph $G = (V_1,V_2,E)$ with $|V_1|=|V_2|=n$ and the goal is
to sample u.a.r from the set $\P$ of perfect and near-perfect matchings of
$G$ (a near-perfect matching is a matching that saturates all but two
vertices of $G$).  Jerrum and Sinclair~\cite{JS-permanent} proposed
the following natural Markov chain $\M_{\rm JS}$ for sampling from
$\P$: At each state $M$, the chain moves to a state $M'$ defined by
the following random experiment:
\begin{itemize}
\item[(i)] Pick $r \in \{0,1\}$ u.a.r and an edge $e \in E$ u.a.r.
\item[(ii)] If $r= 0$ set $M' = M$; Else
\item[(iii)] If $M$ is a perfect matching: Then set $M' = M
\setminus \{e\}$ if $e \in M$, or else $M'=M$.
\item[(iv)] Suppose $M$ is a near-perfect matching. Let $e
=(u,v)$. There are two cases:
\begin{itemize}
\item[(a)] If $u,v$ are both unmatched in $M$, set $M' = M \cup
\{e\}$. {\sf [Add Move]}
\item[(b)] If exactly one of $u,v$ is unmatched, then set $M' = M
\setminus \{e'\} \cup \{e\}$ where $e'$ is the edge in $M$ incident on
whichever of $u,v$ is matched. {\sf [Swap Move]}
\end{itemize}
\item[(v)] If none of the above conditions are met, set $M'=M$.
\end{itemize}

\medskip \noindent {\bf A special graph $G$.} Anil Kumar and
Ramesh~\cite{KR99} show that for a certain graph $G$, every Coupling
strategy on the above chain will require time exponential in $n$. This
graph has some special properties which are used in the
proof; these are:
\begin{enumerate}
\item $G$ has $\Omega(\frac{n!}{c^n})$
perfect matchings for some constant $c > 1$.
\item Each vertex of $G$ has degree at least $\alpha n$, for some
$\alpha < 1/2$.
\item For every pair of vertices, the intersection of their
neighborhoods has size at most $\alpha n/2$.
\end{enumerate}
Such a graph $G$ can be shown to exist using the probabilistic method
(see for example the final version of \cite{KR99}). 

\medskip \noindent {\bf Modeling the Coupling Process.} The coupling
process ${\cal C} = ({\cal X},{\cal Y})$ is specified by transition
probabilities $p_{\cal C}(v,w)$ where $v =(a,b) \in \P \times \P$, and
$w = (c,d) \in \P \times \P$ are pairs of states in $\P$. Note that
$p_{\cal C}(v,w)$ could even be a function of the history, i.e., the
transition probabilities could vary with time (we do not show the time
dependence for notational convenience, but it should be treated as
implicit). Since we are aiming for a negative result and wish to rule
out the existence of {\em any} Coupling based proof, the only thing we will
(and can) assume about these probabilities is that the processes
${\cal X}$ and ${\cal Y}$ must individually be faithful copies of
$\M_{\rm JS}$, or in other words: If $v=(x,y)$, then for each $x' \in
\P$ and for each time instant $t$, $\sum_{w \in T(x')} p_{\cal C}(v,w)
= P(x,x')$ where $T(x')= \{(x',z) ~ | ~ z \in \P\}$ and
$P(\cdot,\cdot)$ is the transition probabilities of the chain $\M_{\rm
JS}$, and a similar equation for $P(y,y')$ for each $y' \in \P$.

\medskip \noindent {\bf Idea behind the Proof.} The basic structure of
the proof is the following: Define a ``distance'' between the two
states $X,Y$ in a Coupling, relative to which the states will have a
tendency to {\em drift away} from each other in {\em any} Coupling,
i.e., most transitions of any Coupling are {\em distance
increasing}. Then analyze this drifting behavior and show that staring
with two states $(X_0,Y_0)$ at a distance $\Theta(n)$ apart, any
Coupling will require exponential number of steps $t$ before the
states $X_t,Y_t$ become equal, with say a probability of $1/2$. This
gives an exponential lower bound on the Coupling time for any
strategy, as desired.

\subsection{Details of the Analysis} 

We partition the states of the Coupling chain ${\cal C}$ into layers
$L(i)$, $i=0,\dots,2n$ according to the ``distance'' $i$ between its
elements, where $L(i)$ contains of all pairs $(M,N) \in \P \times \P$
such that $|M \xor N|=i$. We further partition each set $L(i)$ into
two sets $\bot(i)$ and $\top(i)$, where $\bot(i) = \{(M,N) |~ \exists$
vertex $v$ which is unmatched in exactly one of $M, N\}$, and $\top(i)
= \{(M,N) |$ either both $M$ and $N$ are perfect matchings or both are
near-perfect matchings with the same unmatched vertices$\}$.

A move in ${\cal C}$ from $L(i)$ to $L(j)$ is {\em leftwards} or {\em
distance reducing} if $j < i$, and {\em rightwards} or {\em distance
increasing} if $j > i$. Since $G$ has $\Omega(\frac{n!}{c^n})$ perfect
matchings, with overwhelming probability, the start state of the
Coupling lies in $L(i)$ for some $i \ge n/4$. For simplicity
therefore, we assume that the Coupling ${\cal C}$ begins at some state
in $L(i_0)$, $i_0 \ge n/4$.

The idea now is to upper bound the probabilities of the leftward
transitions and lower bound the probabilities of the rightward
transitions, and then use these bounds to show that the Coupling has a
tendency to drift towards the right. Finally, this will imply that the
(expected) number of steps to reach a state in $L(0)$ will be
exponentially large, giving us our desired result. 

\medskip \noindent {\bf The Key Lemmas.} We now state the main Lemmas
which bound transition probabilities between different layers. We will
later use the statements of these Lemmas give us the desired
``rightward drift''. We give a representative proof of one of the
Lemmas (the proofs of the other Lemmas can be found in \cite{KR99},
and we do not reproduce them here).

\begin{lemma}
\label{lem:coup-cond-1}
No transition in ${\cal C}$ can change the distance by more than $4$.
\end{lemma}

\begin{lemma}
\label{lem:coup-cond-2}
For any coupling strategy, the sum of transition probabilities from
$(M,N) \in \bot(i)$ to vertices in $L(j)$, $j < i$, is at most
$\frac{2i+1}{m}$. 
\end{lemma}

\begin{lemma}
\label{lem:coup-cond-3}
For any coupling strategy, the sum of the transition probabilities
from $(M,N) \in \bot(i)$ into $\bigcup_{j=i+1}^{i+4} \bot(j)$ is at least
$\frac{\alpha n/2 - i - 2}{2m}$. 
\end{lemma}

\begin{lemma}
\label{lem:coup-cond-4}
For any coupling strategy, the sum of the transition probabilities
from $(M,N) \in \bot(i)$ into $\top(i) \cup \top(i+1)$ is at most
$\frac{i+3}{2m}$. 
\end{lemma}

\begin{lemma}
\label{lem:coup-cond-5}
For any coupling strategy, all transitions from $(M,N) \in \top(i)$
are to vertices in either $\top(i)$ or in $\bot(j)$ for some $j \ge
i-2$.
\end{lemma}
\noindent We only prove Lemma~\ref{lem:coup-cond-3} as it is the key Lemma that
establishes a tendency of any Coupling to drift to the right. This
should give a flavor of the sort of arguments necessary to prove the
other Lemmas as well. 

\medskip \noindent {\bf Proof of Lemma~\ref{lem:coup-cond-3}:} Since
$(M,N) \in \bot(i)$, three cases arise: (a) $M$ is a near-perfect
matching and $N$ is a perfect matching; (b) $M$ is a perfect matching
and $N$ is a near-perfect matching; and (c) Both $M$ and $N$ are
near-perfect matchings with at most one common unmatched vertex. Case
(b) is symmetric to Case (a), so we consider Cases (a) and (c) in
turn.

\medskip \noindent {\sf Case (a):} $M$ is near-perfect and $N$ is
perfect. Let $a \in V_1$ and $b \in V_2$ be the unmatched vertices in
$M$. We consider only one situation that will increase $|M \xor N|$
and then lower bound the probability that this situation occurs. The
situation is: $M$ moves to $M' = M + e - (u,u')$ where $e = (a,u)$ and
$(u,u') \in M \cap N$. Now $|M' \xor N| = |M \xor N| + 2$. $N$ can
move to $N'$ where either $N' = N$ or $N' = N - f$ for some edge $f
\in N$. In either case $|M' \xor N'| \ge |M \xor N| + 1$. Furthermore,
$u'$ and $b$ are unmatched in $M'$, and since $(u',b) \notin N$, at
least one of them is matched in $N'$. We thus conclude $(M',N') \in
\bot(j)$, for some $j > i$. Now the probability that this situation
occurs is clearly at least $\frac{\alpha n - | M \setminus N|}{2m}$
which is at least $\frac{\alpha n - i}{2m}$, for any coupling
strategy.

\medskip \noindent {\sf Case (b):} $M$ and $N$ are both
near-perfect. Suppose $M$ have vertices $a \in V_1$ and $b \in V_2$
unmatched and $N$ has vertices $c \in V_1$ and $d \in V_2$
unmatched. Let us assume that $b \neq d$ (while $c$ could equal $a$). 

We once again focus on a particular class of moves which $M$ makes. Suppose
$M$ chooses an edge $e=(b,u)$, where $u$ is {\em not adjacent} to $d$
and $(u,u') \in M \cap N$ for some $u'\in V_2$ (by our assumption
about $G$ there exist at least $\alpha n/2 - |M \setminus N| \ge \alpha n/2 -
i$ such edges $e$. If $e$ is picked (i.e., $M'  = M + e - (u,u')$)
then $|M' \xor N| = |M \xor N| + 2$. It is easy to verify now that the
only moves for $N$ that can reduce the distance back by $2$ are when
it choose the unique edge $(c,c') \in M$, if any, or the unique edge
$(d,d') \in M$, to swap in. The probability of either of these
happening is at most $\frac{2}{2m}$ for any coupling
strategy. Furthermore, in this case $u' \in V_2$ is unmatched in $M'$
and must be matched in $N'$ because $(u,u') \in N$ (it lies in $M \cap
N$) and $(u,d) \notin E$ by the choice of $u$. Hence $(M',N')$ lies in
$\bot()$. Summing up, $(M',N') \in \bot(j)$ for $j > i$ 
with probability at least $\frac{\alpha n/2 -i - 2}{2m}$.   {\hfill
$\Box$ {\em (Lemma~\ref{lem:coup-cond-3})}}

\subsection{Bounding the Coupling Time} 

With the above Lemmas in place, we are ready to finish off the
analysis bounding the coupling time. The rightward drifting behavior
of any Coupling ${\cal C}$ can be predicted (qualitatively) given the
above Lemmas. We now see how to quantify this intuition. We define a
sequence of random variables $Z_0,Z_1,\dots$ which represent the layer
number of some intermediate states of the Coupling. We will show that
$\pr [ Z_t = 0 ] \sim t e^{-\Theta(n)}$, and this will imply an
exponential lower bound on the Coupling time.

Define $Z_0$ to the layer number of the starting state of the Coupling
${\cal C}$. As discussed earlier, we assume $Z_0 \ge n/4$. Also
assume, by virtue of Lemma~\ref{lem:coup-cond-5}, that the starting
state is in a $\bot()$ set rather than a $\top()$ set.

For $i > 0$, the random variable $Z_i$ is defined as follows. If
$Z_{i-1} =0$ then $Z_i = 0$. Otherwise, $Z_i$ is the layer number of
the first state $A$ reached in the Coupling ${\cal C}$ that has the
following properties:
\begin{enumerate}
\item $A \notin L(Z_{i-1})$.
\item $A$ is in some $\bot()$ set or in $L(0)$. 
\end{enumerate}

\begin{lemma}
\label{lem:mart-1}
For every $i \ge 1$, $|Z_i - Z_{i-1}| \le 8$
\end{lemma}
\proof Follows easily from Lemmas~\ref{lem:coup-cond-1} and
\ref{lem:coup-cond-5}. {\hfill $\Box$}

\medskip \noindent The Lemma below quantifies the ``rightward drifting''
behavior of the sequence $Z_0,Z_1,\dots$.
\begin{lemma}
\label{lem:mart-2}
Define $p_i = \frac{\alpha n/2 - Z_{i-1} - 2}{2m}$ and $q_i =
\frac{5(Z_{i-1}+1)}{2m}$. Then $\pr [ Z_i > Z_{i-1} | Z_{i-1} ] \ge
\frac{p_i}{p_i+q_i}$.
\end{lemma}
\proof By Lemma~\ref{lem:coup-cond-3}, $Z_i > Z_{i-1}$ happens with
probability at least $p_i$. By Lemma~\ref{lem:coup-cond-5}, $Z_i <
Z_{i-1}$ only if the first vertex visited after leaving
$\bot(Z_{i-1})$ for the last time is either in $L(j)$, $j < i$, or is
in $\top(Z_{i-1}) \cup \top(Z_{i-1}+1)$. By Lemmas
\ref{lem:coup-cond-2} and \ref{lem:coup-cond-4}, this probability is
at most $q_i$. The claimed result now follows. {\hfill $\Box$}

\medskip Let $\beta > 0$ be a constant such that $\frac{5(\beta n
+1)}{\alpha n/2 - \beta n - 2} \le \frac{1}{16}$. Then it is easy to
see using the above Lemma that
\begin{equation}
\label{eq:coup-cond-1}
\av_{Z_i} ~[ Z_i - Z_{i-1} ~ | ~ Z_{i-1}; 0 < Z_{i-1} \le \beta n ]
\ge \frac{1}{4} \ .
\end{equation}
Combining Lemma~\ref{lem:mart-1} with the above Equation, we will be
able to bound the Coupling time by appealing to the following
submartingale inequality~\cite{KR99} (see also \cite{hajek}). 

\begin{prop}
\label{prop:submart}
Let $Z_0,Z_1,Z_2,\cdots$ be a sequence of random variables with the
following properties (for some $R, \Delta, M > 0$):
\begin{enumerate}
\item $Z_i \ge 0$, for all $i \ge 0$. Further $Z_i = 0$ $\Rightarrow$
$Z_{i+1} = 0$, for all $i \ge 0$.
\item $|Z_i - Z_{i-1}| \le \Delta$ for all $i$, $i \ge 1$.
\item $\av ~ [ Z_i - Z_{i-1} ~ | ~ Z_{i-1}; 0 < Z_{i-1} \le R ] \ge
M$, for all $i$, $i\ge 1$.
\end{enumerate}
Let $T$ be the random variable defined as $\min \{i \ge 0 | Z_i
=0\}$. Then 
\[ \pr [ T \le t | Z_0 ] \le e^{-\frac{MZ_0}{\Delta^2}} + t
e^{-\frac{M(R-\Delta)}{\Delta^2}} \ . \]
\end{prop}

\noindent Note that the above is very similar in spirit to Azuma's
inequality applied to submartingales, except that the assumption (3)
above is made only when conditioned on $0 < Z_{i-1} \le R$, and not
for any value of $Z_{i-1}$ (as is done in Azuma's inequality).

Let us now apply the above Proposition to our setting. Let $t_{\eps}$ be
the earliest instant at which the probability that coupling time
exceeds $t_{\eps}$ falls below $\eps$. Define $T = \min \{i \ge 0 |
Z_i = 0\}$. Then, applying
Proposition~\ref{prop:submart} with $\Delta = 8$, $M = 1/4$, $R =
\beta n$ and $Z_0 \ge n/4$, we get
\[ 1-\eps \le \pr [ T \le t_{\eps}| Z_0 ] \le t_{\eps} e^{-\Theta(n)}
\ . \]
It follows that $t_{\eps} \ge (1-\eps) {\rm exp}(\Theta(n))$. We have
thus proved the following:

\begin{theorem}[\cite{KR99}]
\label{thm:KR99}
Consider any Coupling process for the Markov chain $\M_{\rm JS}$ for
sampling from perfect and near-perfect matchings. The probability that
this process has ``coupled'' exceeds $(1-\eps)$ only after time
$\Omega((1-\eps) e^{\Theta(n)})$. Thus, no proof of rapid mixing of
$\M_{\rm JS}$ exists based on the Coupling Lemma.
\end{theorem}

\section{Concluding Remarks and Open Questions}
\label{sec:conclusions}

We have seen that the mixing rate of a Markov chain is captured by the
spectral gap and also by a geometric parameter called Conductance. We
discussed ways to bound the conductance, and also ways to bound the
spectral gap directly, based on construction of canonical paths or
flows between every pair of states that do not overload any transition
of the Markov chain. The ``flow'' based approach led to the notion of
resistance which also {\em captures} the spectral gap (up to square
factors). We showed that for a large class of chains, the existence of
``good'' canonical paths with low edge-congestion also captures mixing
time, and thus is no weaker than the resistance based approach. We
nevertheless demonstrated that spreading the flow along multiple paths
might still be a very useful design tool by discussing the recent
result of \cite{MS99} on the rapid mixing of a natural chain for sampling
$0$-$1$ knapsack solutions. 

We then turned to an entirely different approach to proving rapid
mixing: Coupling. We discussed ``Path Coupling'' which is a useful
tool in designing good Couplings. We saw several simple and elegant
applications of Coupling which invariably gave much better bounds on
mixing time than known through conductance. One of these examples was
the $0$-$1$ knapsack problem with uniform item sizes for which we
proved a much better mixing time bound than seems possible using the
(more difficult) approach of \cite{MS99}. 

Despite the appeal of Coupling in several applications, it turns out
that Coupling is weaker than conductance in the sense that there are
Markov chains with an exponential gap between their actual mixing time
and that which can be deduced using any Coupling strategy. We
discussed the result of \cite{KR99} which showed such a result for the
famous Jerrum-Sinclair chain for sampling uniformly from the set of
perfect and near-perfect matchings of a bipartite graph.

There are several natural questions on the relative power of the
various techniques that are worthy of more detailed study. We list
some of them below.
\begin{itemize}
\item The result of Kumar and Ramesh~\cite{KR99} is quite natural and
says that Coupling cannot work when there is a measure of distance
relative to which the states have a tendency to drift away from each
other in any Coupling strategy. It will be nice to find other chains
for which Coupling cannot prove rapid mixing. This might shed some
light on how to tackle the question we raise next.
\item Is there a subclass of Markov chains for which Coupling
characterizes rapid mixing (up to polynomial factors)?  What kinds of
structure in the underlying problem enables easy design of good
couplings, i.e., what makes a problem ``Coupling friendly''?
\item It almost seems that whenever Path Coupling works there is a
``natural'' notion of adjacency and a distance metric fixing which
gives a rather easy proof of rapid mixing. For several problems for
which the natural choice for these notions does not work, no known Coupling
based proof seems to be in sight as well. It will be interesting to shed some
light on this, and investigate how one may make Coupling work
when most natural choices for doing Path Coupling do not work out.
\item Finally there are several questions still open about designing
and analyzing rapidly mixing Markov chains for specific sampling problems. Some
of our favorite ones are:
\begin{itemize}
\item Bipartite graphs with a given degree sequence (for sampling
regular bipartite graphs, a rapidly mixing Markov Chain was given
in \cite{KTV97}). More generally, contingency tables with given row
and column sums (the $2 \times n$ case was solved in
\cite{DG-contingency} using Path Coupling).
\item Independent sets in graphs with maximum degree $5$. (The case
$\Delta \le 4$ has been considered in \cite{LV97,DG-indset}, and a
``negative'' result for $\Delta \ge 6$ appears in \cite{DFJ-indset}.)
\item Proper $k$-colorings of a graph when $k < \frac{11}{6} \Delta$.
\item Perfect matchings in a general bipartite graph.
\end{itemize}

\end{itemize}


\begin{thebibliography}{99}

\bibitem{aldous1} D. Aldous. Random walks on finite groups and rapidly
mixing Markov chains. {\em S\'{e}minnaire de Probabilit\'{e}s XVII
1981/82}, Springer Lecture Notes in Mathematics 986, 1983,
pp. 243-297.

\bibitem{aldous} D. Aldous. Some inequalities for reversible Markov
chains. {\em Journal of the London Mathematical Society}, 25 (1982),
pp. 564-576.

\bibitem{alon86} N. Alon. Eigenvalues and expanders. {\em
Combinatorica}, 6 (1986), pp. 83-96.

\bibitem{broder} A. Broder. How hard is it to marry at random? (On the
approximation of the permanent). {\em Proc. of 18th STOC}, pp. 50-58,
1986. 

\bibitem{BD97} R. Bubley and M. Dyer. Path coupling: a technique for
proving rapid mixing in Markov chains. {\em Proc. of 38th FOCS},
pp. 223-231, 1997.

\bibitem{BD-linear-extensions} R. Bubley and M. Dyer. Faster random
generation of linear extensions. {\em Proc. of the 9th ACM Symposium
on Discrete Algorithms}, pp. 35-354, 1998.

\bibitem{BDJ-convex} R. Bubley, M. Dyer and M. Jerrum. An elementary
analysis of a procedure for sampling points in a convex body. {\em
Random Structures and Algorithms}, 12 (1998), pp. 213-235.

\bibitem{diaconis} P. Diaconis and D. Strook. Geometric bounds for
eigenvalues of Markov chains. {\em Annals of Applied Probability}, 1
(1991), pp. 36-61.

\bibitem{DF-volume} M. Dyer and A. Frieze. Computing the volume of
convex bodies: a case where randomness provably helps. In
Probabilistic Combinatorics and its Applications, {\em Proc. of AMS
Symposia in Applied Mathematics}, 44 (1991), pp. 123-170.

\bibitem{DFJ-indset} M. Dyer, A. Frieze and M. Jerrum. On counting
independent sets in sparse graphs. {\em Proc. of 40th FOCS},
pp. 210-217, 1999.

\bibitem{DFK91} M. Dyer, A. Frieze and R. Kannan. A random polynomial
time algorithm for approximating the volume of convex bodies. {\em
Journal of the ACM}, 38 (1991), pp. 1-17.

\bibitem{DFKKPV} M. Dyer, A. Frieze, R. Kannan, A. Kapoor,
L. Perkovic and U. Vazirani. A sub-exponential time algorithm for
approximating the number of solutions to a multidimensional knapsack
problem. {\em Combinatorics, Probability and Computing}, 2 (1993),
pp. 271-284.

\bibitem{DG-contingency} M. Dyer and C. Greenhill. A genuinely
polynomial-time algorithm for sampling two-rowed contingency
tables. {\em Proc. of the 25th International Colloquium on Automata,
Languages and Programming (ICALP)}, pp. 339-350, Aalborg, Denmark
(1998),

\bibitem{DG-coloring} M. Dyer and C. Greenhill, A more rapidly mixing
Markov chain for graph colourings. {\em Random Structures and
Algorithms}, 13 (1998), pp. 285-317.

\bibitem{DG-indset} M. Dyer and C. Greenhill. On Markov chains for
independent sets. {\em Journal of Algorithms}, 35 (2000), pp. 17 - 49.

\bibitem{hajek} B. Hajek. Hitting time and occupation time bounds
implied by drift analysis with applications. {\em Advances in Applied
Probability}, 14 (1982), pp. 502-525.

\bibitem{FM92} T. Feder and M. Mihail. Balanced Matroids. {\em
Proc. of 24th STOC}, pp. 26-38, 1992.

\bibitem{jerrum-coloring} M. Jerrum. A very simple algorithm for
estimating the number of k-colourings of a low-degree graph. {\em
Random Structures and Algorithms}, 7 (1995), pp. 157-165.

\bibitem{jerrum-survey} M. Jerrum. Mathematical foundations of the
Markov chain Monte Carlo method. In {\em Probabilistic Methods for
Algorithmic Discrete Mathematics}, Algorithms and Combinatorics 16,
Springer-Verlag, 1998, pp. 116-165.

\bibitem{JS-permanent} M. Jerrum and A. Sinclair. Approximating the
permanent. {\em SIAM Journal on Computing}, 18 (1989), pp. 1149-1178.

\bibitem{JS-ising} M. Jerrum and A. Sinclair. Polynomial-time
approximation algorithms for the Ising model. {\em SIAM Journal on
Computing}, 22 (1993), pp. 1087-1116.

\bibitem{JS-survey} M. Jerrum and A. Sinclair. The Markov chain Monte
Carlo method: an approach to approximate counting and integration. In
{\em Approximation Algorithms for NP-hard problems}, D.S Hochbaum ed.,
PWS Publishing, Boston, 1997, pp. 482-520.

\bibitem{JVV} M. Jerrum, L. G. Valiant and V. V. Vazirani. Random
generation of combinatorial structures from a uniform
distribution. {\em Theoretical Computer Science}, 43 (1986),
pp. 169-188.

\bibitem{kahale} N. Kahale. {\em A semidefinite bound for mixing rates
of Markov chains}. DIMACS Technical Report 95-41, September 1995.

\bibitem{KTV97} R. Kannan, P. Tetali and S. Vempala. Simple Markov
chain algorithms for generating bipartite graphs and tournaments. {\em
Proc. of 8th SODA}, 1997.

\bibitem{KK90} A. Karzanov and L. Khachiyan. {\em On the conductance
of order Markov chains}, Technical Report DCS 268, Rutgers University,
June 1990.

\bibitem{KR99} V.S. Anil Kumar and H. Ramesh. Coupling vs. conductance
for the Jerrum-Sinclair chain. {\em Proc. of 40th FOCS}, pp. 241-251,
1999.

\bibitem{LR88} T. Leighton and S. Rao. An approximate max-flow min-cut
theorem for uniform multicommodity flow problems with applications to
approximation algorithms. {\em Proc. of 29th STOC}, pp. 422-431, 1988.

\bibitem{LV97} M. Luby and E. Vigoda. Approximately counting up to
four. {\em Proc. of 29th STOC}, pp.  682-687, 1997.

\bibitem{mat91} P. Matthews. Generating random linear extensions of a
partial order. {\em The Annals of Probability}, 19 (1991),
pp. 1367-1392.

\bibitem{mihail89} M. Mihail. Conductance and convergence of Markov
chains: a combinatorial treatment of expanders. {\em Proc. of the 30th
FOCS}, pp. 526-531, 1989.

\bibitem{mihail-broder} M. Mihail. On coupling and the approximation
of the permanent. {\em Information Processing Letters}, 30 (1989),
pp. 91-95.

\bibitem{mohar} B. Mohar. Isoperimetric numbers of graphs. {\em
Journal of Combinatorial Theory, Series B}, 47 (1989), pp. 274-291.

\bibitem{MS99} B. Morris and A. Sinclair. Random walks on truncated
cubes and sampling $0$-$1$ knapsack solutions. {\em Proc. of 40th
FOCS}, pp. 230-240, 1999.

\bibitem{raghavan} P. Raghavan and C. D. Thompson. Randomized
rounding: a technique for provably good algorithms and algorithmic
proofs. {\em Combinatorica}, 7 (1987), pp. 365-374.

\bibitem{sinclair-thesis} A. Sinclair. {\em Algorithms for random
generation and counting: a Markov chain approach}. Ph.D thesis,
University of Edinburgh, June 1988.

\bibitem{sinclair92} A. Sinclair. Improved bounds for mixing rates of
Markov chains and multicommodity flow. {\em Combinatorics,
Probability and Computing}, 1 (1992), pp. 351-370.

\bibitem{SJ89} A. Sinclair and M. Jerrum. Approximate counting,
uniform generation, and rapidly mixing Markov chains. {\em Information
and Computation}, 82 (1989), pp. 93-133.

\bibitem{vadhan} S. Vadhan. Rapidly mixing Markov chains and their
applications. Essay, {\em Churchill College, Cambridge University},
May 1996.

\bibitem{vigoda} E. Vigoda. Improved bounds for sampling
colorings. {\em Proc. of 40th FOCS}, pp. 51-59, 1999.

\end{thebibliography}
\end{document}